\documentclass[conference]{IEEE/IEEEtran}
\IEEEoverridecommandlockouts
\usepackage{algorithm}
\usepackage{algpseudocode}
\usepackage{array}
\usepackage{tabularx}

\usepackage{cite}
\usepackage{amsmath,amssymb,amsfonts}
\usepackage{graphicx}
\usepackage{textcomp}
\usepackage{xcolor}
\usepackage{soul}
\usepackage{listings}
\usepackage[normalem]{ulem}
\usepackage[T1]{fontenc} 
\usepackage{csvsimple} % Required package for handling CSVs
\usepackage{orcidlink}

\def\BibTeX{{\rm B\kern-.05em{\sc i\kern-.025em b}\kern-.08em
    T\kern-.1667em\lower.7ex\hbox{E}\kern-.125emX}}
\begin{document}

\title{Memory-Access Profiling for Early Multicore Interference Awareness in Embedded Systems}

%IEEE AUTHORS INFO
\author{\IEEEauthorblockN{1\textsuperscript{st} Jean-Baptiste Lefoul\orcidlink{0000-0002-5984-8319}}
\IEEEauthorblockA{\textit{Computer Engineering and Software Engineering} \\
\textit{Polytechnique Montreal}\\
Montreal, Canada \\
jean-baptiste.lefoul@polymtl.ca}
\and
\IEEEauthorblockN{2\textsuperscript{nd} Ahmad Shahnejat Bushehri\orcidlink{0000-0001-8254-7242}}
\IEEEauthorblockA{\textit{Computer Engineering and Software Engineering} \\
\textit{Polytechnique Montreal}\\
Montreal, Canada }
\and
\IEEEauthorblockN{3\textsuperscript{rd} Alaric Chan Lock\orcidlink{0009-0003-3878-236X}}
\IEEEauthorblockA{\textit{Computer Engineering and Software Engineering} \\
\textit{Polytechnique Montreal}\\
Montreal, Canada}
\and
\IEEEauthorblockN{4\textsuperscript{th} Felipe Gohring de Magalhaes\orcidlink{0000-0002-0766-1421}}
\IEEEauthorblockA{\textit{Computer Engineering and Software Engineering} \\
\textit{Polytechnique Montreal}\\
Montreal, Canada }
\and
\IEEEauthorblockN{5\textsuperscript{th} Gabriela Nicolescu\orcidlink{0000-0002-5205-9931}}
\IEEEauthorblockA{\textit{Computer Engineering and Software Engineering} \\
\textit{Polytechnique Montreal}\\
Montreal, Canada }
}

\maketitle

\begin{abstract}
Applications executing on different multicore processors may interfere through shared resources. Such interference can introduce execution-time variability and complicate system integration, especially when applications that behave predictably in isolation are co-scheduled on the same multicore platform.

This paper presents a lightweight profiling-based framework for early multicore interference awareness. The proposed approach captures source-level memory-operation traces from C applications with limited source-code modification and combines them with architectural parameters of the target platform. The framework maps memory accesses to cache sets and memory blocks, generates cache-occupancy profiles and heatmaps, and evaluates pairwise interference risks through a progressive three-stage analysis. First, a coarse-grained screening step classifies applications according to their non-stack memory-access behavior. Second, a static multicore cache-profiling step identifies cache sets whose combined footprint exceeds or approaches the hardware associativity. Third, a temporal footprint analysis tracks intervening unique blocks between consecutive accesses to the same memory block and computes Contention Potential Growth to identify execution windows vulnerable to cache eviction.

The framework is evaluated using four benchmark pairings executed on a dual-core platform, with hardware performance counters measuring L2 cache misses used as validation references. The results show that the proposed analysis is consistent with the observed high- and low-risk cases and provides conservative warnings for near-capacity configurations. In particular, different applications pairs might produces up to 213.09\% increase in L2 cache misses. These results demonstrate that memory-access profiling can provide practical and interpretable early indicators of cache-related multicore interference before exhaustive hardware testing.
\end{abstract}


\section{Introduction}

The increasing computational demands of modern embedded systems have accelerated the adoption of multicore processors. By allowing several applications or tasks to execute in parallel, multicore architectures provide higher throughput, better resource utilization, and improved energy efficiency compared with traditional single-core platforms. These advantages make multicore processors attractive for domains such as avionics, automotive systems, industrial control, telecommunications, and high-performance embedded computing.

However, multicore execution introduces timing and performance challenges that are largely absent from single-core systems. Even when applications are logically independent and assigned to different cores, they may still interact through shared hardware resources such as last-level caches, interconnects, memory controllers, and main memory. As a result, the execution behavior of one application may be affected by the activity of another application running on a different core. These shared-resource effects are commonly referred to as multicore interference.

Multicore interference can occur in both direct and indirect forms. Direct interference occurs when several cores request access to the same shared resource during overlapping time intervals, such as a shared bus, memory controller, or memory bank. Indirect interference occurs when the activity of one core modifies the hardware state observed by another core. Shared-cache eviction is a typical example: one application may evict cache lines that another application would otherwise reuse, thereby increasing cache misses and execution time. Both forms of interference can reduce timing predictability and complicate the integration of independently developed applications.

This problem is particularly important for performance-critical and safety-conscious embedded systems. Applications that exhibit stable behavior when executed in isolation may experience significant slowdowns when co-executed with other workloads. Consequently, system designers need methods that can identify potentially harmful application pairings and shared-resource hotspots before final deployment on the target platform.

Several families of techniques have been proposed to analyze or mitigate multicore interference. Static WCET and timing-analysis methods can provide conservative timing bounds, but they often require detailed architectural models and may become pessimistic on complex multicore processors~\cite{10.1145/1347375.1347389}. Analytical cache-interference models can reason about shared-cache effects, but they usually depend on assumptions about replacement policies, timing models, or access-arrival patterns~\cite{10.1145/3575757.3593643}. Dynamic measurement-based approaches, including hardware performance-counter analysis and stress testing, capture real hardware effects but typically require executing workload combinations on the target platform and often reveal interference only after it has occurred~\cite{9113125}. Mitigation techniques such as cache partitioning, memory-bandwidth regulation, and interference-aware memory allocation can reduce interference, but they require enforcement mechanisms, hardware support, or optimization procedures~\cite{6531079,9113100}. Locality-based approaches such as reuse-distance and stack-distance profiling provide useful information about memory behavior, but they may require specialized profiling mechanisms or detailed trace processing~\cite{5388318,10.1007/978-3-031-06156-1_49}.

These existing approaches leave a practical gap for lightweight and interpretable interference-awareness methods during early development. At this stage, designers may not yet require a formal WCET bound or a cycle-accurate simulation model. Nevertheless, they still need to determine whether a given pair of applications is likely to stress shared cache or memory resources. They also need interpretable information about where interference may originate, including which cache sets are overloaded, which access patterns are problematic, and during which execution windows cache-eviction risk becomes significant.

This paper addresses this need by proposing a profiling-based framework for multicore interference awareness. The proposed framework analyzes the memory-access behavior of C applications and combines the resulting traces with architectural parameters of the target multicore platform. Rather than attempting to reproduce every microarchitectural event, the framework provides a hardware-aware approximation of how applications use cache sets, memory blocks, and access patterns. The objective is not to replace formal timing analysis or cycle-accurate simulation, but to provide an early and interpretable indication of potential interference risks.

The proposed methodology follows a progressive three-stage analysis. First, each application is classified according to the proportion of non-stack memory accesses. This coarse-grained screening step quickly identifies application pairs that are unlikely to generate significant memory-path interference. Second, when at least one application exhibits significant non-stack memory activity, the framework merges the traces and computes joint cache-set occupancy and heatmap profiles. This spatial analysis identifies cache sets whose combined unique-block footprint exceeds or approaches the physical cache associativity. Third, for suspicious regions, the framework performs temporal footprint analysis by tracking the intervening unique footprint between consecutive accesses to the same memory block. This analysis computes Contention Potential Growth ($\Delta CP$), which quantifies how much the co-running application expands eviction-vulnerable execution windows.

The framework is validated on an NXP QorIQ P2020RDB-PCA dual-core platform using bare-metal applications. Four application pairings are evaluated: Dijkstra versus Floyd-Warshall, FFT versus Jacobi-2D, AES-128 ECB versus Conv2D, and Bsort versus ADPCM Encoder. Hardware performance counters measuring total L2 data cache misses are collected during isolated and co-running executions and are used as validation references. The results show that the framework identifies the Dijkstra/Floyd-Warshall pair as high risk, which is confirmed by a 213.09\% increase in L2 cache misses. It also identifies the Bsort/ADPCM pair as low risk, with negligible L2 miss variation. For intermediate cases, the framework provides conservative or refined medium-risk assessments, demonstrating its usefulness as an early warning mechanism.

The main contributions of this paper are summarized as follows:
\begin{itemize}
    \item A lightweight memory-access profiling methodology for C applications with limited source-code modification.
    \item A hardware-aware mapping framework that projects memory traces onto cache sets and memory blocks.
    \item A progressive interference-risk analysis combining non-stack access screening, joint cache occupancy profiling, and temporal footprint analysis.
    \item A Contention Potential Growth ($\Delta CP$) metric that quantifies the increase in eviction-vulnerable windows caused by co-running applications.
    \item A validation study on an NXP QorIQ P2020RDB-PCA multicore platform comparing predicted risk classes with hardware-measured L2 cache-miss variations.
\end{itemize}

The remainder of this paper is organized as follows. Section~\ref{background_section} introduces the basic concepts of shared hardware resources, cache behavior, memory banks, stack and non-stack memory accesses, and direct and indirect interference channels. Section~\ref{SOTA_section} reviews related work on WCET analysis, multicore interference, hardware mitigation, empirical stress testing, and locality-based estimation. Section~\ref{method_section} presents the proposed methodology, including cache-footprint analysis, intervening unique footprint analysis, and interference-risk assessment. Section~\ref{results_section} reports the validation results and hardware measurements. Section~\ref{discussion_section} discusses the interpretation of the results, limitations, and future work. Finally, Section~\ref{conclusion_section} concludes the paper.
\section{Basic Concepts}\label{background_section}

This section introduces the architectural and memory-behavior concepts used throughout the paper. The focus is on shared hardware resources in multicore systems, cache mapping and eviction, bus and interconnect contention, memory-bank behavior, stack and non-stack memory accesses, and the general role of memory-access behavior in multicore interference.

\subsection{Shared Hardware Resources and Interference Channels}
\label{subsec:shared_resources_interference}

Modern multicore processors improve computational throughput by allowing multiple applications or tasks to execute simultaneously on separate processing cores \cite{10915487}. Although each core can execute its own instruction stream independently, several hardware resources remain shared across the platform. These shared resources commonly include last-level caches, interconnects, memory controllers, and main memory banks \cite{10.1145/3687308}.

\begin{figure}[!h]
    \centering
    \includegraphics[width=0.6\linewidth]{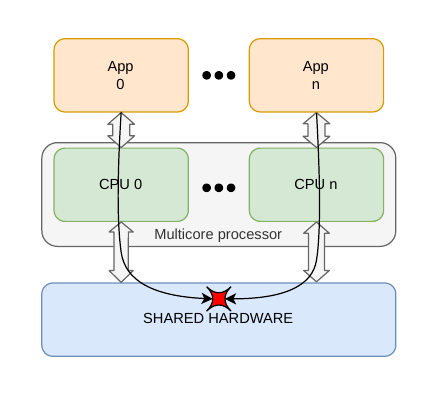}
    \caption{Multicore interference in shared resources}
    \label{fig:Interference}
\end{figure}

Sharing hardware resources introduces dependencies between otherwise independent applications \cite{11257230}, as shown in Fig.~\ref{fig:Interference}. Even if two applications do not communicate at the software level, they may still interact through the memory hierarchy. For example, one application may increase the latency of another application by occupying shared cache space, generating frequent memory requests, or repeatedly accessing the same memory bank. As a result, the execution behavior of one application may influence the performance observed by another.

Multicore interference can be categorized into direct and indirect forms \cite{11257197}. Direct interference occurs when multiple applications simultaneously request access to the same shared resource. Typical examples include contention for a shared bus, a memory controller, a shared cache, or a memory bank. Since these resources can serve only a limited number of requests at a given time, simultaneous demand may increase latency and reduce throughput.

Indirect interference occurs when one application modifies the hardware state observed by another application. Shared-cache eviction is a common example \cite{kou2024cache}. An application that heavily utilizes a cache set may evict cache lines belonging to another application, forcing subsequent accesses by the affected application to be served from lower levels of the memory hierarchy. Unlike direct interference, indirect interference may remain visible even after the interfering request has completed, because the shared hardware state has already been modified.

Both forms of interference can increase execution-time variability and complicate multicore integration \cite{meunier:tel-04457889}. Direct interference is primarily associated with simultaneous resource demand, whereas indirect interference is mainly associated with changes in the shared hardware state. Therefore, an interference-awareness approach should consider both the spatial distribution of memory accesses and their temporal overlap during execution.

\subsection{Cache Behavior}
\label{subsec:cache_behavior}

Caches are small and fast memories placed between the processor and main memory \cite{10.1145/3583139}. Their purpose is to reduce average memory-access latency by storing recently or frequently used data and instructions \cite{10.1145/3508042}. When the processor requests a memory address, the cache is checked first. If the requested block is present, the access is served quickly from the cache. Otherwise, the data must be fetched from a lower level of the memory hierarchy, which increases latency.

\begin{table}[!t]
\centering
\scriptsize
\caption{Address decomposition for cache mapping.}
\label{tab:address_decomposition}
\renewcommand{\arraystretch}{1.7}
\setlength{\tabcolsep}{2pt}
\begin{tabularx}{\columnwidth}{|
>{\hsize=1.50\hsize\centering\arraybackslash}X|
>{\hsize=1.03125\hsize\centering\arraybackslash}X|
>{\hsize=0.46875\hsize\centering\arraybackslash}X|}
\hline
\textbf{TAG} &
\textbf{INDEX} &
\textbf{OFFSET} \\
\textbf{(Bits 31--16)} &
\textbf{(Bits 15--5)} &
\textbf{(Bits 4--0)} \\
\hline
\end{tabularx}
\end{table}

A cache is usually organized as a set of cache lines \cite{10773698}. Each cache line stores a block of contiguous memory bytes. In a set-associative cache, the cache is divided into several sets, and each set contains several ways \cite{10384476}. A memory block maps to exactly one cache set, but it may be stored in any way of that set \cite{9739802}. If the cache has $S$ sets, $W$ ways, and a cache-line size of $B$ bytes, then the total cache capacity \cite{correia2024runtime} is:

\begin{equation}
C = S \times W \times B .
\label{eq:cache_capacity}
\end{equation}

A memory address can be decomposed into three fields \cite{11119683}: offset, index, and tag, as illustrated in Table \ref{tab:address_decomposition} for a cache with 32 ($=2^5$) bytes line size and 2048 ($=2^{11}$) sets. The offset identifies the byte position inside a cache line. The index identifies the cache set to which the memory block maps. The tag is used to determine whether the requested memory block is actually stored in the selected set.

Given a memory address $a$, a cache-line size $B$, and a cache with $S$ sets, the cache-line number $L(a)$ and cache set $\mathrm{Set}(a)$ are computed as:

\begin{equation}
L(a) = \left\lfloor \frac{a}{B} \right\rfloor ,
\label{eq:cache_line_number}
\end{equation}

\begin{equation}
\mathrm{Set}(a) = L(a) \bmod S .
\label{eq:cache_set}
\end{equation}

The corresponding tag is computed as:

\begin{equation}
\mathrm{Tag}(a) = \left\lfloor \frac{L(a)}{S} \right\rfloor .
\label{eq:cache_tag}
\end{equation}

When a memory request is issued, the cache uses the index bits to select the corresponding set. Then, the tag stored in each way of that set is compared with the tag of the requested address. If one valid way contains a matching tag, the access is a cache hit. If no valid way contains the requested tag, the access is a cache miss. A cache miss requires fetching the requested block from a lower memory level, such as a lower-level cache or main memory.

Cache misses can occur for several reasons \cite{curto2023improving}. A compulsory miss occurs when a memory block is accessed for the first time. A capacity miss occurs when the cache cannot contain the application's full working set. A conflict miss occurs when several memory blocks map to the same cache set and compete for the limited number of ways in that set. Conflict misses are particularly important in multicore interference analysis because independent applications may repeatedly access addresses that map to the same shared-cache sets.

When a cache miss occurs and the target set is full, the cache must evict an existing cache line to make room for the new block. The choice of the evicted line is determined by the cache replacement policy, such as Least Recently Used (LRU), pseudo-LRU, random replacement, or First-In First-Out (FIFO) \cite{10.1145/3634737.3637677}. Eviction is a central mechanism in indirect interference \cite{10400823}. If one core fills a shared-cache set with its own data, it may evict useful cache lines belonging to another core. When the affected core later accesses the evicted blocks, additional cache misses occur, increasing execution time.

Cache write policies also influence memory traffic \cite{benedicte_illescas_2022}. In a write-through cache, every write operation is immediately propagated to the next memory level. This simplifies consistency but may increase bus and memory traffic. In a write-back cache, modified data are written to lower memory levels only when the corresponding cache line is evicted. This reduces write traffic but introduces dirty cache lines, which may cause additional latency upon eviction. Therefore, the write policy affects both local cache behavior and shared-resource usage.

Another relevant design choice is the allocation policy on write misses \cite{benedicte_illescas_2022}. With a write-allocate policy, a missed write causes the corresponding block to be loaded into the cache before being modified. With a no-write-allocate policy, writes bypass the cache and are sent directly to a lower memory level. Write-allocate policies may improve locality when written data are reused, but they may also increase cache pressure. No-write-allocate policies reduce cache pollution but may increase direct memory traffic.

In multicore systems, cache behavior becomes more complex because one core's cache activity may affect another core's performance. Shared caches are especially sensitive to cross-core evictions \cite{10609602}. If two applications frequently access addresses mapping to the same cache sets, they may repeatedly evict each other's cache lines. This effect can occur even when the applications are completely independent at the software level.

Cache-coherency mechanisms may also introduce additional traffic in systems where multiple cores access shared writable data \cite{11257375}. Coherency protocols ensure that all cores observe a consistent view of memory, often by invalidating or updating cache lines across cores \cite{9773254}. However, when the analyzed applications are assumed to be independent and do not intentionally share writable data, coherency-induced interference is not the primary focus. In that case, cache-set contention and cache-line eviction remain the dominant cache-related effects considered by the proposed analysis.

\subsection{Bus and Interconnect Behavior}
\label{subsec:bus_interconnect}

The bus or interconnect transports requests and data between processing cores, caches, memory controllers, and peripheral components. In multicore systems, the interconnect is a shared communication medium. Therefore, simultaneous memory requests generated by different cores may compete for access to it.

A key element of bus and interconnect behavior is arbitration \cite{11336342}. When multiple requesters attempt to use the interconnect simultaneously, an arbitration mechanism determines which request is served first. Common arbitration policies include fixed-priority arbitration \cite{11645615}, round-robin arbitration \cite{10673654}, time-division arbitration \cite{11452392}, and quality-of-service-aware arbitration \cite{11335930}. The selected policy directly affects latency, fairness, and predictability.

Bus contention occurs when multiple cores issue memory requests that overlap in time. Since the interconnect has finite bandwidth, requests may be queued or delayed. This contention is a form of direct interference because it results from simultaneous demand for the same shared resource. Applications with high memory-request rates are more likely to generate bus contention, especially when they frequently miss in the cache and access lower memory levels.

Another important factor is the frequency disparity between processor cores and shared interconnects. Processor cores often operate at higher frequencies than memory buses and external memory interfaces \cite{10504070}. Therefore, a processor can generate memory requests faster than the memory subsystem can serve them. If several cores generate such requests concurrently, the interconnect and memory controller can become bottlenecks.

The type of memory command also affects bus efficiency. Alternating read and write transactions may introduce turnaround overhead because the bus or memory controller may require additional cycles to switch direction between read and write modes. Applications that frequently alternate between reads and writes may therefore create more interconnect overhead than applications with more regular access patterns.

In the context of interference analysis, bus behavior is relevant because it links cache misses and memory accesses to execution-time degradation. A program that frequently misses in the cache not only incurs local miss penalties but also increases pressure on the shared interconnect. When several applications behave similarly, the combined pressure may cause performance degradation.

\subsection{Memory Banks}
\label{subsec:memory_banks}

Main memory is commonly organized into banks to increase parallelism and throughput \cite{10196422}. A memory bank can be viewed as a semi-independent memory region that can serve requests independently of other banks, depending on the memory controller design \cite{10.1145/3716368.3735168}. By distributing accesses across multiple banks, the system can exploit bank-level parallelism and improve memory bandwidth.

Memory-bank mapping determines which physical addresses are assigned to which memory banks. Depending on the address-mapping policy, consecutive memory addresses may be distributed across multiple banks or concentrated within a single bank. This mapping strongly influences interference \cite{9812711}. If two applications frequently access addresses mapped to different banks, their requests may experience less contention. If they frequently access the same bank, they may compete for the same internal memory resource.

A bank conflict occurs when multiple requests target the same memory bank within overlapping time intervals. Since a bank can generally serve only a limited number of requests at a time, bank conflicts increase memory-access latency. This type of interference is particularly important for memory-intensive applications whose working sets do not remain in cache.

Memory-bank behavior is also affected by row-buffer locality \cite{11605627}. Modern dynamic memories organize each bank into rows. When a row is activated, subsequent accesses to the same row can be served more quickly than accesses requiring a different row. Therefore, sequential or localized memory accesses may benefit from row-buffer hits, whereas irregular access patterns may cause frequent row activations and precharges.

Although detailed row-buffer and memory-controller modeling may be beyond the scope of a lightweight profiling framework, identifying bank concentration remains useful for estimating memory-level interference risks \cite{11320188}. From an interference-awareness perspective, memory-bank analysis helps determine whether multiple applications concentrate their memory accesses on the same physical memory regions. Such concentration increases the probability of bank conflicts and memory controller contention.

\subsection{Stack and Non-Stack Memory Accesses}
\label{subsec:stack_non_stack}

The proposed framework uses the ratio of stack-to-non-stack memory accesses as a coarse-grained screening criterion. Stack accesses typically correspond to local variables, function call frames, return addresses, and temporary values.~\cite{compArch7th} In many embedded applications, stack data exhibit relatively localized behavior because stack frames are allocated and released in a structured manner following the program's call hierarchy~\cite{8726496}.

Non-stack accesses include heap-allocated data, global variables, static data, arrays, buffers, and other memory regions that may have wider spatial distribution~\cite{compArch7th}. Such accesses are often more likely to populate multiple cache sets and create larger working sets~\cite{8726496}. Therefore, applications with a high proportion of non-stack accesses are more likely to generate cache pressure, memory-path contention, or bank conflicts~\cite{Olson2016Revisiting}.

\subsection{Memory-Access Behavior as an Interference Indicator}
\label{subsec:memory_access_indicator}

Most interference mechanisms originate from memory activity~\cite{6382387}. Applications continuously generate instruction fetches, data reads, data writes, cache accesses, and memory transactions~\cite{compArch7th}. These operations determine how the application uses the memory hierarchy and how much pressure it places on shared resources.

Consequently, memory-access traces provide valuable information for estimating an application's susceptibility to interference and its likelihood of interfering with other workloads~\cite{10.1007/978-3-319-10214-6_4}. An application with frequent memory requests, poor locality, concentrated cache-set usage, or repeated accesses to the same memory regions is more likely to be sensitive to co-execution effects~\cite{10.1145/2188286.2188294, 10.1145/3575757.3593643}.

Rather than directly simulating every hardware transaction, interference awareness can be supported by characterizing memory behavior through representative indicators~\cite{7120141}. Examples include memory-access frequency, stack and non-stack access distribution, cache-set concentration, access locality, and memory-region usage~\cite{10.1145/3695794.3695817,jolivel:hal-05448514}. These indicators provide an initial view of whether an application is likely to create pressure on shared resources.

These concepts do not replace cycle-accurate simulation or hardware measurements. Instead, they motivate the profiling-based methodology introduced in Section~\ref{method_section}.  %optional
\section{State-of-the-Art}\label{SOTA_section}

This section reviews prior work related to multicore interference analysis and positions the proposed framework with respect to existing approaches. The literature can be grouped into four complementary families. The first family focuses on hard real-time timing analysis and Worst-Case Execution Time (WCET) estimation. The second family studies hardware-level contention mechanisms and mitigation techniques for shared resources. The third family relies on empirical benchmarking, stress testing, and hardware monitoring to expose interference effects on real platforms. The fourth family develops analytical, statistical, and locality-based models to estimate interference from memory-access behavior, cache footprint, and reuse properties.

\subsection{Hard Real-Time and WCET Analysis Frameworks}
\label{subsec:sota_wcet}

WCET analysis aims to compute or estimate an upper bound on the execution time of a task. In hard real-time systems, this bound is required to demonstrate deadline satisfaction. However, modern processors complicate this problem because execution time depends not only on the control-flow path of the program, but also on caches, pipelines, branch predictors, memory hierarchy state, and shared-resource contention. The general WCET problem and its classical decomposition into control-flow analysis, processor-behavior analysis, and bound calculation are extensively surveyed in~\cite{10.1145/1347375.1347389}.

\subsubsection{Static Analytical Approaches}
\label{subsubsec:sota_static_wcet}

Static WCET analysis derives timing bounds without executing the program. It typically operates on source code, intermediate representation, or binary code, reconstructs a control-flow graph, derives loop bounds and infeasible-path constraints, models the target architecture, and computes a global timing bound using methods such as implicit path enumeration. Static methods are attractive because they can provide conservative bounds, which are necessary for certification-oriented systems. Abstract interpretation is a central technique in this family because it can safely over-approximate possible program states and hardware states~\cite{10.1145/1347375.1347389}.

In multicore systems, however, static analysis becomes significantly more difficult. A task's execution time may be affected by tasks running on other cores through shared caches, buses, and main memory. For shared caches, static approaches must bound whether accesses from one core can evict cache blocks used by another core. Authors in~\cite{10.1145/3575757.3593643} proposed an analysis of shared-cache interference using event-arrival curves, where cache accesses issued by competing cores are treated as event streams. Their method derives tighter bounds on inter-core cache interference by considering timing information and the maximum rate at which interfering accesses can arrive~\cite{10.1145/3575757.3593643}. This work shows that analytical reasoning about shared-cache interference is possible, but it still requires detailed cache assumptions, such as the replacement policy, and remains oriented toward safe WCET bounding rather than lightweight early-stage profiling.

Static approaches therefore provide strong guarantees, but their applicability depends on the availability of detailed architectural models and analyzable software structures. When hardware components are undocumented, when cache replacement or memory-controller policies are complex, or when dynamic memory behavior dominates, purely static methods may become overly pessimistic or difficult to deploy. In contrast, the proposed framework does not attempt to derive a safe timing bound. It provides an early diagnostic layer that highlights workload pairings and cache regions that should be prioritized for deeper analysis or hardware validation.

\subsubsection{Dynamic Measurement-Based Approaches}
\label{subsubsec:sota_dynamic_measurement}

Measurement-based approaches estimate timing behavior by executing the program and collecting execution times or hardware events. The simplest form is end-to-end measurement, where the whole program is executed for a set of inputs and the maximum observed time is used as an estimate. Such results are realistic for the observed executions, but they are generally not safe because the worst-case input, worst-case path, and worst-case hardware state may not have been exercised~\cite{10.1145/1347375.1347389}.

Authors in~\cite{deverge_et_al:OASIcs.WCET.2005.808} investigated how measurement-based WCET estimation could be made safer by combining structural testing, program clustering, and control of hardware sources of timing variability~\cite{deverge_et_al:OASIcs.WCET.2005.808}. Their approach decomposes the program into smaller segments, generates test data to cover program paths within those segments, and uses compiler and hardware-control techniques to reduce timing variability. This work is important because it highlights two key issues also relevant to interference analysis: measurements must be sufficiently representative, and microarchitectural state must be controlled or characterized to avoid misleading conclusions.

Dynamic approaches are also widely used in multicore interference analysis through hardware monitoring. Performance Monitoring Counters (PMCs) can record cache misses, stall cycles, memory transactions, and execution cycles while applications execute alone or with co-runners~\cite{9113125}. Such counters provide valuable evidence of real interference effects, but they are platform-dependent and typically reveal interference after it has occurred. They also do not automatically explain which memory regions, cache sets, or access patterns caused the observed degradation.

Software-based tracing and instrumentation frameworks provide another way to collect runtime behavior. Dynamic analysis and whole-system timing exploration have been used to improve measurement coverage and expose timing-sensitive executions~\cite{11315096}. Dynamic Binary Instrumentation (DBI) can modify executable code at the instruction level to inject probes and analysis routines. Recent work such as NOProbe proposes a lightweight user-space DBI framework based on binary rewriting, compiler-generated NOPs, pseudo-NOPs, and safe concurrent patching to reduce probe insertion and execution overhead~\cite{11283063}. Other tracepoint mechanisms also aim to provide flexible runtime observability with reduced instrumentation cost~\cite{https://doi.org/10.1002/spe.2746}. Such techniques are relevant because they demonstrate how runtime behavior can be collected with limited overhead. However, these works primarily focus on instrumentation and tracing mechanisms, whereas the present work focuses on using memory-access profiles to estimate multicore interference risks.

\subsubsection{Hybrid Verification Methodologies}
\label{subsubsec:sota_hybrid_wcet}

Hybrid WCET analysis combines static program reasoning with timing measurements~\cite{10.1145/3807211.3807227}. Instead of relying entirely on detailed hardware models, hybrid methods measure or estimate the execution time of smaller code regions and then use static path analysis to compose a global timing estimate. This reduces the need for complete microarchitectural modeling while improving coverage compared with pure end-to-end measurements.

TimeWeaver is a representative hybrid WCET tool that combines static value and path analysis with timing measurements obtained from non-intrusive instruction-level real-time traces~\cite{kastner_et_al:OASIcs.WCET.2019.1}. By relying on hardware-supported tracing rather than software instrumentation, it avoids the probe effect associated with instrumentation code that modifies the original execution and may perturb cache or pipeline behavior~\cite{11315096,11283063,https://doi.org/10.1002/spe.2746}. The tool uses trace segments to construct worst-case timing estimates and also provides information about trace coverage. This is particularly relevant for modern multicore processors where exact timing models may be unavailable or unrealistic.

Recent work has also introduced machine-learning-based hybrid timing estimation. Authors in~\cite{amalou:hal-04148587} uses Transformer-XL to estimate the timing of basic blocks while static techniques identify the longest execution path. This approach is motivated by the difficulty of manually modeling complex processors. By learning context-aware timing behavior from measured basic blocks, this framework reduces dependence on explicit microarchitectural models. However, such methods require training data and focus on timing estimation rather than resource-specific interference localization.

Nowotsch et al. proposed an interference-sensitive WCET methodology that combines offline analysis with runtime resource-capacity enforcement~\cite{6932594}. Their approach separates core-local timing from shared-resource interference delays and uses runtime monitoring to enforce resource-usage assumptions. This direction is important because it demonstrates how analysis and enforcement can be combined to control multicore interference. In contrast, the framework proposed in this paper does not enforce resource budgets at runtime; instead, it estimates interference risk from memory-access profiles to support early integration decisions.

\subsection{Multicore Resource Contention and Hardware Mitigation}
\label{subsec:sota_resource_contention}

A second body of work focuses on the hardware mechanisms that create multicore interference and on architectural or system-level techniques that reduce their impact. These studies are essential because interference is not a single phenomenon: it may arise from shared interconnects, memory controllers, DRAM banks, shared caches, or operating-system memory-management decisions.

\subsubsection{Shared Interconnects and Bus Arbitration}
\label{subsubsec:sota_bus}

Shared buses and interconnects are significant sources of direct interference. When several cores request access to shared memory or shared cache levels, the interconnect must arbitrate between them. Arbitration policies directly affect worst-case latency and timing predictability.

Kelter et al. studied bus-aware multicore WCET analysis for TDMA-arbitrated buses~\cite{6001765}. In TDMA arbitration, each core receives a fixed time slot. This makes the worst-case bus access delay bounded, but simple bounds can be highly pessimistic because they assume the worst possible alignment between memory requests and TDMA slots. Their work derives tighter bounds on bus-access durations by considering TDMA offset information, thereby improving WCET precision.

Bourgade et al. proposed predictable bus-arbitration schemes for heterogeneous time-critical workloads~\cite{6059179}. Instead of assigning the same bandwidth to all cores, their two-level arbitration groups cores and provides different bandwidth levels while preserving computable worst-case latencies. This is relevant for systems where tasks have different memory-bandwidth demands. These bus-oriented approaches show that predictability can be improved through arbitration design, but they require hardware-level assumptions or arbitration policies that may not exist in a given commercial off-the-shelf platform.

\subsubsection{Main Memory and DRAM Controllers}
\label{subsubsec:sota_dram}

Main memory is another major interference source. Modern DRAM systems are organized into channels, ranks, banks, rows, and columns. Requests to different banks can often proceed in parallel, whereas requests to the same bank or to different rows in the same bank may serialize and incur additional latency.

Kim et al. explicitly modeled the major resources in commercial off-the-shelf DRAM systems, including banks, buses, and memory-controller behavior, to bound memory interference delay~\cite{6925998}. Their work is important because it shows that treating memory as a black-box resource with a constant service time can be unsafe or inaccurate. Modern memory controllers may reorder requests, and DRAM access latency depends on row-buffer state, bank conflicts, and command scheduling.

Fernández de Lecea Navarro studied multicore contention in memory controllers and DRAM using performance monitoring and event-monitoring mechanisms~\cite{fernandez_de_lecea_navarro_2024}. This work is relevant because it focuses on tracking contention at the level of the memory controller and main memory, which complements approaches that only observe application-level slowdown.

More recent work by Sullivan et al. argues that DRAM interference is fundamentally a bank-level phenomenon and proposes per-bank memory-bandwidth regulation~\cite{11558396}. Their analysis shows that aggregate bandwidth regulation can be overly pessimistic because it ignores DRAM bank-level parallelism. By regulating bandwidth at the bank level, their approach preserves temporal isolation while improving throughput. This observation supports the broader idea that interference analysis should account not only for the total number of memory accesses, but also for their distribution across hardware resources.

\subsubsection{Cache-Level Interference and Isolation}
\label{subsubsec:sota_cache}

Caches are among the most studied sources of multicore unpredictability. In shared last-level caches, tasks running on different cores can evict each other's cache lines, producing inter-core interference. Gracioli et al. surveyed cache-management mechanisms for real-time embedded systems and classified cache interference into intratask, intracore, and intercore interference~\cite{10.1145/2830555}. They also reviewed mitigation strategies such as cache partitioning, cache locking, memory allocators, compiler support, and page coloring.

Mancuso et al. proposed Single Core Equivalence (SCE), an OS-level framework that attempts to export a single-core-like abstraction from a multicore platform~\cite{7176036}. SCE combines mechanisms such as cache partitioning, DRAM bank partitioning, and memory-bandwidth regulation to isolate cores and allow per-core analysis. This approach is highly relevant to certification-oriented multicore systems because it seeks to preserve modular reasoning. However, SCE is a mitigation and isolation framework, whereas the proposed work aims to identify interference risk before such mitigation decisions are applied.

Cache-mitigation techniques reduce interference by controlling cache allocation. However, they often require hardware support, page coloring, cache-way allocation, or explicit memory-layout control. They also require prior knowledge about which tasks are cache-sensitive and which cache sets or ways are likely to be contested. The proposed framework contributes at this earlier diagnostic stage by deriving cache-set occupancy, heatmap, and temporal footprint indicators from memory-access traces.

\subsubsection{Operating-System and Memory-Management Interventions}
\label{subsubsec:sota_os_memory}

Operating-system-level mechanisms can mitigate interference by regulating how cores access shared memory resources \cite{9842219}\cite{torresauroradugo:hal-03697093}. MemGuard is a representative memory-bandwidth reservation system that separates guaranteed and best-effort memory bandwidth~\cite{6531079}. It uses per-core bandwidth budgets to provide temporal isolation while reclaiming unused bandwidth to improve throughput. Such approaches demonstrate that software can help control memory interference on commercial off-the-shelf platforms.

Yao et al. extended this direction with schedulability analysis for memory-bandwidth-regulated multicore real-time systems~\cite{7093140}. Their model regulates each core's memory access rate and derives worst-case response-time bounds based on the task under analysis, the assigned bandwidth, and the number of cores. This provides modularity because the analysis does not require detailed knowledge of tasks running on other cores.

Memory management can also be addressed at compile time. Reder and Becker proposed an interference-aware memory-allocation approach for real-time multicore systems~\cite{9113100}. Their method formulates the allocation of program variables to distributed memory segments as an Integer Linear Programming (ILP) problem and uses a generic platform model that can represent buses, routers, memories, and NoC-based interconnects. Unlike runtime bandwidth regulation, this technique attempts to reduce interference by placing data in memory regions that minimize conflicting access routes. This is relevant to the present work because both approaches use memory-access information to reason about interference. However, their objective is optimization of memory placement for WCET reduction, whereas the proposed framework focuses on profiling-based interference estimation and reporting.

These OS-level and memory-management approaches improve predictability by enforcing or optimizing resource usage. In contrast, the proposed work does not throttle applications, partition resources, or solve an allocation problem. Instead, it provides a profiling-based estimation of which application pairs and cache regions are likely to create interference. This information can be used before selecting mitigation strategies such as memory-bandwidth regulation, cache partitioning, memory allocation, or core allocation.

\subsection{Empirical Contentious Benchmarking and Stress Testing}
\label{subsec:sota_empirical}

A third research direction evaluates multicore platforms by intentionally creating contention and measuring the resulting degradation. This family of work is particularly relevant because it provides physical evidence of interference on real hardware.

\subsubsection{Synthetic Resource Enemies}
\label{subsubsec:sota_enemies}

Synthetic resource enemies are micro-benchmarks engineered to saturate specific shared resources, such as cache sets, miss-status holding registers, write-back buffers, memory buses, or DRAM banks. By co-running a target application with such an enemy program, one can estimate how much slowdown the target may suffer under aggressive contention.

Iorga et al. studied how to measure and tune multicore interference using enemy programs~\cite{9113125}. Their work highlights that empirical interference results can vary significantly depending on measurement methodology, operating-system control, frequency scaling, thread migration, and statistical treatment. They propose a more reliable measurement methodology based on percentiles and confidence intervals, and they use auto-tuning techniques to generate architecture-specific enemy programs.

This line of work is valuable because it reveals real contention effects that may not be visible in static models. However, enemy-based testing is inherently experimental. It requires executing many workload and stressor combinations and may not explain why a specific application is vulnerable except through post-hoc analysis. The proposed framework complements this direction by estimating likely cache-interference mechanisms from application memory profiles and then validating these estimates through hardware measurements.

\subsubsection{Observational Measurement Frameworks}
\label{subsubsec:sota_observational}

Observational frameworks use hardware counters, traces, or runtime monitors to record degradation during active tests. PMCs can provide measurements such as cache misses, memory transactions, execution cycles, branch events, and stall cycles. These measurements are useful for comparing isolated and interfered executions.

Nowotsch et al. used runtime resource monitoring and capacity enforcement to support interference-sensitive WCET analysis~\cite{6932594}. Their approach demonstrates that runtime observation can be combined with offline reasoning to bound shared-resource interference. TimeWeaver also uses non-intrusive real-time traces to derive timing estimates while avoiding the probe effect associated with software instrumentation~\cite{kastner_et_al:OASIcs.WCET.2019.1}. Memory-controller-oriented monitoring approaches, such as~\cite{fernandez_de_lecea_navarro_2024}, further show the relevance of PMCs and event monitors for observing contention in the memory subsystem.

The proposed work adopts the same principle that hardware measurements are necessary for validation, but uses them differently. Hardware counters are not the primary source of the interference estimate. Instead, the estimate is derived from memory-access profiling, and hardware counters are used as reference evidence to evaluate whether the estimated risk classes are coherent with observed cache-miss variation.

\subsection{Analytical Interference Estimation and Locality Metrics}
\label{subsec:sota_locality}

A fourth family of work develops mathematical, statistical, or locality-based models to predict interference from program or workload characteristics. These methods are directly related to the proposed framework because they seek to explain performance degradation using compact representations of memory behavior.

\subsubsection{Statistical and Global Footprint Models}
\label{subsubsec:sota_statistical}

Several models predict cross-core performance degradation using aggregate pressure on shared resources. Zhao et al. proposed a regression-based approach in which performance degradation is modeled as a piecewise function of aggregate pressures on shared resources, such as shared cache and memory bandwidth~\cite{7120141}. Their results show that an application's degradation can be predicted from the total pressure generated by co-located workloads, rather than from the exact identity of those workloads.

Cache-sharing models similarly attempt to predict the behavior of combined workloads from individual workload profiles. Babka et al. studied the accuracy of cache-sharing models based on stack-distance profiles and emphasized that model accuracy depends on the quality of profile inputs, workload stability, replacement-policy assumptions, and experimental control~\cite{10.1145/2188286.2188294}. This observation is highly relevant to the proposed work: lightweight metrics can provide useful interference awareness, but they must be interpreted as approximations whose validity depends on trace representativeness and hardware assumptions.

\subsubsection{Time-Averaged and Window-Based Profiling Metrics}
\label{subsubsec:sota_window_metrics}

Time-averaged and window-based profiling metrics summarize application behavior over an interval. Examples include memory-access rate, cache-miss rate, bandwidth demand, stall ratio, or per-window access distribution. Yao et al. used memory-intensity and stall-ratio concepts to derive response-time bounds under bandwidth regulation~\cite{7093140}. Fischer and Falk used event-arrival curves to represent the rate at which interfering cache accesses may arrive over time~\cite{10.1145/3575757.3593643}. These approaches show that interference is not only spatial, but also temporal: two applications may map to the same cache set or memory bank, but interference is only relevant if the conflicting accesses occur within overlapping time intervals.

Reder and Becker also emphasize the importance of identifying memory accesses that may occur in parallel~\cite{9113100}. Their synchronization-based interference model determines which code regions may overlap and then estimates the cost of conflicting routes through shared components such as buses, routers, and memories. This is closely related to window-based profiling because both approaches attempt to avoid treating all accesses in a complete execution as if they could interfere simultaneously. The difference is that Reder and Becker use this information to optimize memory allocation and WCET bounds, while the proposed work uses trace-derived indicators to estimate and report interference risk.

This motivates the proposed paper's use of temporal footprint analysis. Whole-trace cache occupancy provides a useful first approximation, but it may overestimate interference when conflicting accesses are separated in time.

\subsubsection{Reuse and Stack Distance Formulations}
\label{subsubsec:sota_reuse_distance}

Reuse distance and stack distance are classical locality metrics. Mattson et al. introduced stack processing as an efficient method for evaluating storage hierarchies from address traces~\cite{5388318}. The central idea is that the distance between two accesses to the same item provides information about whether the second access is likely to hit or miss in a cache of a given capacity.

Modern reuse-distance profiling extends this idea to multicore systems. Sasongko et al. proposed a low-overhead reuse-distance profiling tool for multicore applications using PMUs and debug registers~\cite{10.1007/978-3-031-06156-1_49}. Their work considers private and shared caches and accounts for inter-thread invalidations, showing that reuse distance remains relevant for multicore locality analysis.

These locality metrics provide a deeper view of cache behavior than simple access counts. However, they may require address sampling, debug registers, binary instrumentation, or specialized profiling mechanisms. The proposed framework adopts a simpler source-level memory-operation profiling strategy and derives coarser but interpretable indicators, such as cache-set pressure and intervening unique footprint. This makes the approach less precise than full reuse-distance analysis, but potentially easier to use during early development.

The proposed framework is related to this family because it uses an intervening unique footprint between consecutive accesses to the same memory block. This footprint is conceptually close to stack distance because it estimates how many distinct cache-mapped blocks may displace a block before reuse. However, the proposed method does not aim to provide a full reuse-distance profile or simulate exact cache behavior as developed by authors in \cite{9185451}. Instead, it uses per-cache-set footprint growth as an interpretable indicator of eviction vulnerability under co-execution.

\subsection{Positioning of the Proposed Work}
\label{subsec:sota_positioning}

The reviewed literature shows that multicore interference has been addressed from several angles. Static WCET methods provide conservative timing guarantees but require detailed architectural models and may become pessimistic on complex multicore platforms. Dynamic measurement-based methods capture real behavior but require representative executions and often reveal interference only after co-execution has occurred. Hybrid WCET frameworks combine static path reasoning with measurements, but their objective remains timing-bound estimation. Hardware and OS-level mitigation techniques reduce interference through arbitration, bandwidth regulation, cache partitioning, memory allocation, or memory-bank control, but they require enforcement mechanisms, optimization procedures, or architectural support. Empirical enemy-based testing exposes real interference, but it may require many experiments and does not directly provide source-level explanations. Reuse-distance, stack-distance, and regression-based models can predict aspects of cache behavior or performance degradation, but they often depend on specialized profiles, training data, or assumptions about access distributions.

The proposed framework occupies a different position. It is not intended to provide a formal WCET bound, replace cache simulation, solve memory allocation, or enforce runtime isolation. Instead, it provides an early interference-awareness layer based on memory-access profiling of C applications. The framework maps profiled memory operations to cache sets and memory blocks, generates cache-occupancy profiles and heatmaps, classifies workload pairings using non-stack memory-access behavior, and applies temporal footprint analysis to identify vulnerable execution windows. 

Compared with static WCET approaches, the proposed method sacrifices formal guarantees in favor of lower modeling complexity and higher practical observability. Compared with dynamic hardware measurements, it estimates potential interference before exhaustive multicore testing. Compared with empirical enemy-based testing, it provides more interpretable information about where interference risk originates. Compared with full reuse-distance profiling, it uses a simpler per-cache-set intervening-footprint analysis that is easier to interpret during early development. Compared with interference-aware memory allocation, it does not attempt to optimize placement decisions, but it can provide diagnostic information needed before such optimization is applied.

Therefore, the contribution of this work is best understood as an intermediate layer between analytical timing analysis and empirical validation: a lightweight, hardware-aware, profile-based methodology for identifying interference-prone workload combinations, overloaded cache sets, and execution windows vulnerable to cross-core cache eviction before final deployment.
\section{Methodology}\label{method_section}

\begin{figure}[!t]
    \centering
    \includegraphics[width=0.8\linewidth]{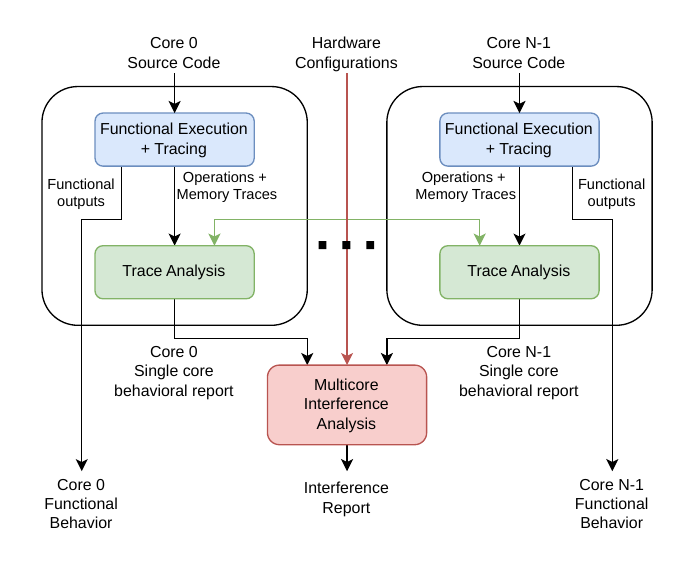}
    \caption{Framework flow.}
    \label{fig:framework}
\end{figure}

The proposed framework estimates multicore interference risk through dynamic memory profiling and hardware-aware trace analysis. The objective is to provide early interference awareness without requiring full cycle-accurate simulation or exhaustive execution of all workload combinations on the target platform. As illustrated in Fig.~\ref{fig:framework}, the framework takes as input the source code of the applications to be analyzed and a description of the relevant architectural parameters of the target hardware. Each application is first executed and profiled independently. The resulting traces are then mapped to cache sets, memory blocks, and memory-hierarchy indicators. Finally, the single-application profiles are combined to estimate the potential interference risk when applications are co-scheduled on different cores.

The methodology consists of six main components: framework input specification, trace generation, trace normalization and hardware mapping, cache-footprint and intervening-footprint analysis, single-application memory characterization, and pairwise multicore interference-risk assessment.

\subsection{Framework Inputs}

The framework requires two categories of inputs: application source code and hardware configuration.

\subsubsection{Application Source Code}

The first input consists of one or more C applications intended to execute on a multicore platform. Each application is profiled independently in order to capture its intrinsic memory-access behavior without interference from other applications. This independent characterization is necessary because the objective is to estimate whether applications that behave normally in isolation may interfere with each other when executed concurrently.

The framework targets applications written in C while minimizing manual source-code modification. Instead of manually inserting instrumentation code around every memory operation, the framework relies on type substitution and profiling wrappers to intercept operations on program variables. Although the analyzed applications are written in C, the profiling executable is generated using a C++ wrapper layer. Standard C types are replaced by profiling template types during profiling execution only. The original application logic is preserved, while the profiling executable is used to collect memory-operation traces.

\subsubsection{Hardware Configuration}

The second input describes the relevant architectural characteristics of the target multicore platform. The required parameters include:
\begin{itemize}
    \item number of processing cores;
    \item cache hierarchy;
    \item cache size;
    \item number of cache sets;
    \item cache associativity;
    \item cache-line size;
    \item memory-bank configuration, when available;
    \item memory-address mapping, when available;
    \item bus or interconnect organization, when available.
\end{itemize}

These parameters allow the framework to map memory operations to hardware resources. For example, cache parameters determine the cache set and memory block associated with each memory address, while memory-bank information can be used to estimate potential bank-level contention. The objective is not to reproduce all microarchitectural behavior cycle by cycle, but to obtain a hardware-aware approximation of how each application uses shared resources.

\subsection{Trace Generation}

The trace-generation phase collects memory-related events during application execution. The traced events include reads, writes, variable accesses, object constructions, dynamic allocations and deallocations, and function-level events. Each event is recorded with sufficient information to support later hardware-aware analysis.

A trace event is represented as:
\begin{equation}
    e_k = \langle app, op, addr, size, obj, seq \rangle ,
    \label{eq:trace_event}
\end{equation}
where $app$ identifies the profiled application, $op$ represents the operation type, $addr$ is the accessed memory address, $size$ is the access size when available, $obj$ identifies the associated program object when known, and $seq$ represents the chronological order of the event in the trace.

The generated trace represents a source-level approximation of memory behavior rather than a cycle-accurate sequence of hardware load and store instructions. The objective is to capture relative memory-access patterns, object usage, access concentration, and resource-mapping tendencies, not to reproduce the exact instruction-level memory stream generated by the final optimized binary.

\subsubsection{Type-Based Profiling Mechanism}

To reduce manual instrumentation, the framework defines a generic template class, denoted as \texttt{ProfilerTypeTemplate<T>}. Standard C types are replaced through preprocessor definitions or controlled type aliases. For example, a variable originally declared as \texttt{uint32\_t} can be replaced by \texttt{ProfilerTypeTemplate<uint32\_t>}. As a result, operations applied to that variable are redirected through the profiling class.

This mechanism allows the framework to intercept standard operations such as assignments, arithmetic operations, comparisons, increments, and compound assignments. Whenever such an operation is executed, the profiling class records the corresponding operation type and memory information before forwarding the operation to the original value.

For example, the original code in Listing~\ref{lst:example_source} contains standard integer variables and arithmetic operations. After type substitution, operations on these variables are automatically routed through the profiling template. Listing~\ref{lst:preprocessed_code} shows the corresponding profiling version after preprocessing. Therefore, the framework can record that \texttt{total\_bugs} was read, modified, and written during execution. Table~\ref{tab:trace_sample} shows the simplified generated trace from executing the example source code.

\begin{lstlisting}[language=C, caption={Example C source code.}, label={lst:example_source}, breaklines=true, basicstyle=\tiny\ttfamily]
#include <stdint.h>

int main(void)
{
    int32_t total_bugs = 10;
    int32_t fixed_bugs = 4;
    uint8_t new_bugs_created = 2;

    total_bugs = total_bugs - fixed_bugs;
    total_bugs += (int32_t)new_bugs_created;

    return total_bugs;
}
\end{lstlisting}

\begin{lstlisting}[language=C++, caption={Preprocessed profiling version of the example source code.}, label={lst:preprocessed_code}, breaklines=true, basicstyle=\tiny\ttfamily]
#include <stdint.h>

int main(void)
{
    ProfilerTypeTemplate<int32_t> total_bugs = 10;
    ProfilerTypeTemplate<int32_t> fixed_bugs = 4;
    ProfilerTypeTemplate<uint8_t> new_bugs_created = 2;

    total_bugs = total_bugs - fixed_bugs;
    total_bugs += (ProfilerTypeTemplate<int32_t>)new_bugs_created;

    return (int)total_bugs;
}
\end{lstlisting}

\begin{table}[!t]
\centering
\caption{Generated access trace from the execution of Listing~\ref{lst:example_source}.}
\label{tab:trace_sample}
\scriptsize
\renewcommand{\arraystretch}{1.1}
\begin{tabular}{cll}
\hline
\textbf{\#} & \textbf{Access Type} & \textbf{Access Address} \\
\hline
0  & CONSTR    & @total\_bugs \\
1  & CONSTR    & @fixed\_bugs \\
2  & CONSTR    & @new\_bugs\_created \\
3  & SUB\_R    & @fixed\_bugs \\
4  & SUB\_R    & @total\_bugs \\
5  & CONSTR    & @temp\_var \\
6  & ASSIGN\_R & @temp\_var \\
7  & ASSIGN\_W & @total\_bugs \\
8  & READ      & @new\_bugs\_created \\
9  & CONSTR    & @temp\_var \\
10 & ADD\_R    & @temp\_var \\
11 & ADD\_W    & @total\_bugs \\
12 & READ      & @total\_bugs \\
13 & CONSTR    & @ret\_val \\
\hline
\end{tabular}
\end{table}

The main advantage of this mechanism is that it captures high-level memory behavior while requiring limited changes to the original program. However, the approach must be carefully implemented for operations involving pointers, arrays, structures, library calls, and explicit memory functions such as \texttt{memcpy} or \texttt{memset}. These cases may require additional wrappers or compiler-assisted instrumentation.

\subsubsection{Functional Output Preservation}

During profiling execution, the framework also records the functional output of each application. This includes return values, standard outputs, and optional application-level state summaries. This information could be used to verify that the profiling mechanism does not modify the expected behavior of the application. Therefore, each profiled execution produces both a functional report and a memory-operation trace.

\subsection{Trace Normalization and Hardware Mapping}

Once traces are collected, raw events are normalized and mapped to the corresponding hardware resources. This step transforms memory-operation traces into cache and memory-hierarchy usage profiles.

\subsubsection{Trace Normalization}

Raw traces may contain auxiliary events introduced by the profiling mechanism, including object construction, temporary variables, type conversions, and wrapper-level operations. Before computing interference indicators, the trace is normalized to retain or label the events relevant to memory-behavior analysis.

The normalization step groups operations belonging to the same source-level expression when needed, distinguishes read and write accesses, identifies stack and non-stack accesses, and associates each event with the corresponding application object when possible. 
%Temporary and wrapper-generated events can either be removed or explicitly labeled so that profiling artifacts are not interpreted as application-level memory pressure.

\subsubsection{Cache Mapping}

For each memory access, the framework maps the accessed address to the corresponding cache line, cache set, and tag using the cache-address decomposition introduced in Section~\ref{subsec:cache_behavior}. In particular, the cache-line number, set index, and tag are computed according to Eq.~\eqref{eq:cache_line_number}, Eq.~\eqref{eq:cache_set}, and Eq.~\eqref{eq:cache_tag}, respectively.

This mapping enables the framework to determine which cache sets are used by each application and how many distinct memory blocks are mapped to each set. The resulting set-level view is then used to compute cache-set occupancy, identify overloaded or near-capacity sets, and detect regions where shared-cache eviction may occur during co-execution.

\subsubsection{Memory-Bank Mapping}

When memory-bank information is available, memory accesses can also be mapped to memory banks according to the target memory configuration. If the platform uses a simple modulo-based bank mapping, the bank associated with address $a$ can be approximated as:
\begin{equation}
    \mathrm{Bank}(a) =
    \left\lfloor \frac{a}{B_m} \right\rfloor \bmod N_b ,
    \label{eq:method_bank_mapping}
\end{equation}
where $B_m$ is the memory-bank interleaving granularity and $N_b$ is the number of memory banks. If the target platform uses a more specific mapping function, the framework replaces this approximation with the platform-specific address-to-bank mapping.

The current validation focuses primarily on shared-cache interference using L2 cache-miss measurements. However, the same hardware-mapping step can be extended to memory banks and memory-controller resources when the necessary architectural information and counters are available.

\subsubsection{Access-Type Mapping}

Each memory operation is also classified according to its access type. The framework distinguishes reads, writes, allocations, deallocations, construction events, standard operations, and function-level events. This classification supports the identification of memory-intensive behavior and helps distinguish computational operations from memory-pressure sources.

% \subsection{Cache Footprint and Intervening Unique Blocks}
\subsection{Overview of the Proposed Cache-Footprint Analysis}
\label{subsec:cache_footprint}

The proposed framework relies on two complementary views of cache usage. The first view is spatial and measures how many distinct memory blocks are mapped to each cache set. This cache-set occupancy analysis identifies sets whose footprint approaches or exceeds the hardware associativity. The second view is temporal and measures the intervening unique footprint between consecutive accesses to the same memory block. This temporal analysis identifies reuse windows in which a block may become vulnerable to eviction.

Spatial occupancy alone is not sufficient because two applications may use the same cache set during different execution phases. Therefore, the framework combines cache-set occupancy with intervening-footprint analysis. The formal definitions of isolated occupancy, joint occupancy, critical vulnerability windows, and Contention Potential Growth ($\Delta CP$) are introduced in the following subsections.

\subsection{Single-Application Memory Characterization}

After hardware mapping, each application is characterized independently. This produces a single-application behavioral report describing how the application uses the memory hierarchy. The report includes cache-set utilization, distinct memory-block usage, stack and non-stack access ratios, heatmap visualizations, and isolated temporal footprint statistics.

\subsubsection{Cache-Set Occupancy}

For each application $A$ and cache set $s$, the framework records the set of distinct memory blocks accessed by the application and mapped to $s$. This set is denoted as:
\begin{equation}
    \mathcal{B}_{A}(s) =
    \{ b \mid b \text{ is accessed by } A \text{ and } \mathrm{Set}(b)=s \}.
    \label{eq:block_set_single}
\end{equation}

The single-application occupancy of set $s$ is then:
\begin{equation}
    O_A(s) = |\mathcal{B}_{A}(s)| .
    \label{eq:single_occupancy}
\end{equation}

If $O_A(s)$ approaches or exceeds the cache associativity $W$, the application may already exert significant pressure on that cache set even when executed alone. This does not necessarily imply multicore interference, but it identifies regions where the application is sensitive to additional cache pressure from a co-running workload.

\subsubsection{Intervening Unique Footprint}
\label{subsec:cache_characterization}

To identify temporal windows where a memory block is vulnerable to eviction, the framework computes an intervening unique footprint for each reused memory block. This footprint is computed per cache set and is conceptually close to stack distance.

\begin{figure}[!t]
    \centering
    \includegraphics[width=.8\linewidth]{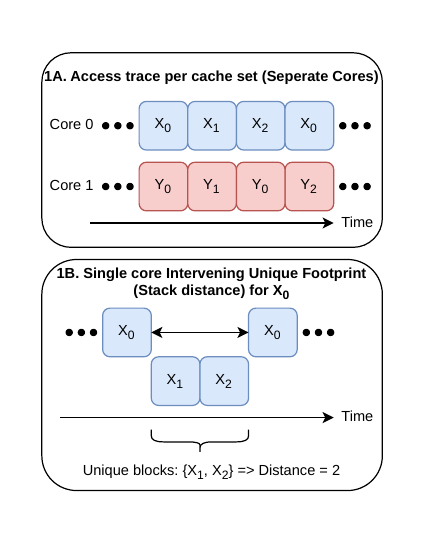}
    \caption{Determining the intervening unique footprint for a memory block using separate single-core traces.}
    \label{fig:Stack_distance_SC}
\end{figure}

Let $b$ be a memory block mapped to cache set $s$. Consider two consecutive accesses to $b$ at trace positions $t_m$ and $t_{m+1}$, as illustrated in Fig.~\ref{fig:Stack_distance_SC}. The intervening unique footprint in isolated execution is defined as the number of distinct blocks mapped to the same cache set that are accessed between these two positions:
\begin{equation}
    I^{solo}_{b,m}
    =
    \left|
    \left\{
    b' \mid
    b' \neq b,\ 
    \mathrm{Set}(b')=\mathrm{Set}(b),\
    t_m < seq(b') < t_{m+1}
    \right\}
    \right| .
    \label{eq:solo_intervening_footprint}
\end{equation}

In an ideal LRU cache, a block becomes vulnerable to eviction when the number of distinct intervening blocks reaches the cache associativity. However, real hardware may use pseudo-LRU or other replacement policies and may evict a block before this theoretical limit is reached. Therefore, the framework uses a configurable critical footprint threshold:
\begin{equation}
    \theta = \beta W ,
    \label{eq:critical_threshold}
\end{equation}
where $W$ is the cache associativity and $\beta \in (0,1]$ is a user-defined conservatism factor. In the default configuration, $\beta=0.75$, which corresponds to a threshold of six unique blocks for an 8-way set-associative cache.

A critical vulnerability window in isolated execution is then defined as:
\begin{equation}
    CW^{solo}_{b,m}
    =
    \begin{cases}
    1, & \text{if } I^{solo}_{b,m} \geq \theta, \\
    0, & \text{otherwise}.
    \end{cases}
    \label{eq:solo_critical_window}
\end{equation}

The isolated critical count is computed as:
\begin{equation}
    CC_{solo}
    =
    \sum_{b}
    \left(
    1 +
    \sum_{m}
    CW^{solo}_{b,m}
    \right).
    \label{eq:cc_solo}
\end{equation}

This value represents a baseline contribution for each tracked memory block, augmented by the number of reuse windows that are already close to the cache-capacity limit during isolated execution.

\subsubsection{Heatmap Generation}

The framework also generates cache heatmaps. In these heatmaps, one axis represents cache sets and the other represents unique memory blocks. The intensity of each cell represents the number of accesses or the occupancy pressure associated with that cache region.

These heatmaps help distinguish between different memory-access patterns. A uniform distribution across many blocks may indicate broad cache pressure, while a concentration on a small number of hot blocks may indicate localized reuse. This distinction is important because high spatial occupancy does not always imply high temporal interference.

\subsubsection{Stack and Non-Stack Access Classification}

The framework computes the ratio of non-stack accesses for each application:
\begin{equation}
    R^{nonstack}_A =
    \frac{N^{nonstack}_A}{N^{stack}_A + N^{nonstack}_A},
    \label{eq:nonstack_ratio}
\end{equation}
where $N^{stack}_A$ and $N^{nonstack}_A$ denote the number of stack and non-stack memory accesses, respectively.

This ratio is used as a coarse-grained screening indicator. Stack-dominated applications often exhibit localized memory behavior, whereas non-stack accesses, including heap, global, static, array, and buffer accesses, may be distributed across wider memory regions. Therefore, applications with a high non-stack ratio are more likely to generate cache-set pressure or memory-path contention.

\subsection{Pairwise Multicore Interference Estimation}

After each application has been characterized independently, the framework evaluates pairwise interference risk. For clarity, the methodology is described for two applications, $A$ and $B$, executing on two cores. The same principle can be extended to more cores by applying the analysis to multiple pairings or by constructing a combined trace containing all co-scheduled applications.

\subsubsection{Trace Interlacing}

To approximate co-execution, the trace of the target application is interlaced with the trace of the co-running application. The resulting merged trace represents a combined access stream in which memory blocks from both applications may compete for the same cache sets. If timestamps are available, the interlacing preserves temporal order. If only event order is available, the framework uses a normalized ordering policy to approximate concurrent execution.

This interlaced trace is not intended to reproduce exact cycle-level scheduling. Instead, it provides a conservative approximation of how the memory footprints of two applications may overlap in shared hardware resources.

\subsubsection{Joint Cache Occupancy}

For each cache set $s$, the framework computes the combined set of distinct blocks accessed by both applications:
\begin{equation}
    \mathcal{B}_{A,B}(s)
    =
    \mathcal{B}_{A}(s) \cup \mathcal{B}_{B}(s).
    \label{eq:joint_block_set}
\end{equation}

The joint occupancy is:
\begin{equation}
    O_{A,B}(s) = |\mathcal{B}_{A,B}(s)| .
    \label{eq:joint_occupancy}
\end{equation}

If $O_{A,B}(s) > W$, the combined footprint exceeds the physical associativity of the cache set and indicates a potential static capacity violation. If $O_{A,B}(s)$ does not exceed $W$ but approaches the critical threshold $\theta$, the pair is classified as a near-capacity case and may require hardware validation or optional temporal analysis.

\subsubsection{Multicore Intervening Unique Footprint}

The intervening unique footprint is recomputed over the interlaced multicore trace. Let $b$ be a memory block belonging to the target application. For two consecutive accesses to $b$ at positions $t_m$ and $t_{m+1}$ in the target trace, the multicore intervening footprint includes distinct blocks from both the target and co-running applications that map to the same cache set.

\begin{figure}[!t]
    \centering
    \includegraphics[width=0.8\linewidth]{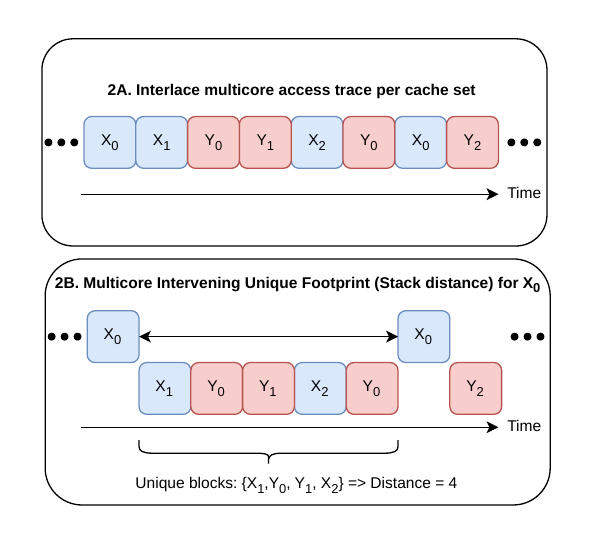}
    \caption{Determining the intervening unique footprint for a memory block in the estimated multicore trace.}
    \label{fig:Stack_distance_MC}
\end{figure}

As illustrated in Fig.~\ref{fig:Stack_distance_MC}, the multicore intervening footprint is defined as:
\begin{equation}
    I^{multi}_{b,m}
    =
    \left|
    \left\{
    b' \mid
    b' \neq b,\ 
    \mathrm{Set}(b')=\mathrm{Set}(b),\
    t_m < seq_{multi}(b') < t_{m+1}
    \right\}
    \right| .
    \label{eq:multi_intervening_footprint}
\end{equation}

The multicore critical vulnerability window is then:
\begin{equation}
    CW^{multi}_{b,m}
    =
    \begin{cases}
    1, & \text{if } I^{multi}_{b,m} \geq \theta, \\
    0, & \text{otherwise}.
    \end{cases}
    \label{eq:multi_critical_window}
\end{equation}

The multicore critical count is:
\begin{equation}
    CC_{multi}
    =
    \sum_{b}
    \left(
    1 +
    \sum_{m}
    CW^{multi}_{b,m}
    \right).
    \label{eq:cc_multi}
\end{equation}

\subsubsection{Contention Potential Growth}

To isolate the additional vulnerability introduced by the co-running application, the framework compares the multicore critical count with the isolated critical count. The Contention Potential Growth is defined as:
\begin{equation}
    \Delta CP =
    \frac{CC_{multi} - CC_{solo}}{CC_{solo}}
    \times 100 .
    \label{eq:delta_cp}
\end{equation}

A $\Delta CP$ value close to zero indicates that the co-running application does not significantly increase the number of vulnerable reuse windows. A high $\Delta CP$ value indicates that the co-running application expands the number of windows in which target memory blocks are likely to be evicted. Therefore, $\Delta CP$ is used as a temporal indicator of cache-interference risk.

This metric should be interpreted as an interference-awareness indicator rather than an exact predictor of execution-time degradation. It identifies cases where the co-running application increases cache-eviction vulnerability, but the final hardware impact may also depend on replacement policy, prefetching, write buffers, memory-controller behavior, and other microarchitectural effects.

\subsection{Interference Risk Assessment}
\label{subsec:risk_assessment}

The proposed analysis framework evaluates potential hardware-level interference between independent applications co-running on a multicore system. Because the evaluated applications are assumed to occupy isolated memory spaces and do not explicitly communicate, any performance degradation is attributed primarily to contention through shared physical resources. The analysis is structured as a progressive screening process, summarized in Algorithm~\ref{alg:multicore_interference}.

\begin{algorithm}[!t]
\caption{Multicore Interference Risk Analysis}
\label{alg:multicore_interference}
\scriptsize
\begin{algorithmic}[1]
\Require Trace files $T_A$ and $T_B$ for applications $A$ and $B$
\Require Stack-access threshold $\tau_{stack}$, cache associativity $W$
\Require Critical threshold factor $\beta$
\Ensure Interference risk profile and vulnerability report

\State $\tau_{stack} \gets 0.10$
\State $\beta \gets 0.75$
\State $\theta \gets \beta W$

\Statex \textbf{Step 1: Coarse-Grained Application Classification}
\State $r_A \gets \textsc{CalculateNonStackAccessRatio}(T_A)$
\State $r_B \gets \textsc{CalculateNonStackAccessRatio}(T_B)$

\If{$r_A < \tau_{stack}$ \textbf{and} $r_B < \tau_{stack}$}
    \State \Return \textsc{Report}(\textit{Low risk: both applications exhibit limited non-stack memory activity.})
\EndIf

\Statex \textbf{Step 2: Static Multicore Cache Profiling}
\State $T_{merged} \gets \textsc{MergeExecutionTraces}(T_A,T_B)$
\State $O_{joint} \gets \textsc{ComputeJointCacheOccupancy}(T_{merged})$
\State $H_{joint} \gets \textsc{ComputeJointCacheHeatmap}(T_{merged})$
\State $O_{max} \gets \max_{s}(O_{joint}(s))$
\State $\mathcal{S}_{problem} \gets \emptyset$

\ForAll{cache set $s \in O_{joint}$}
    \If{$O_{joint}(s) > W$}
        \State $\mathcal{S}_{problem} \gets \mathcal{S}_{problem} \cup \{(s, O_{joint}(s), O_{joint}(s)-W)\}$
    \EndIf
\EndFor

\If{$O_{max} < \theta$}
    \State \Return \textsc{Report}(\textit{Low risk: combined cache footprint remains below the critical threshold.})
\ElsIf{$\mathcal{S}_{problem} = \emptyset$}
    \State \Return \textsc{Report}(\textit{Low-to-medium risk: combined cache footprint approaches associativity but does not exceed it.})
\EndIf

\State \textsc{GenerateStaticInterferenceReport}$(\mathcal{S}_{problem}, H_{joint})$

\Statex \textbf{Step 3: Temporal Footprint Analysis}
\State $\mathcal{C}_A \gets \emptyset$, $\mathcal{C}_B \gets \emptyset$

\ForAll{application $P \in \{A,B\}$}
    % \State $T_P \gets \textsc{SelectTrace}(P,T_A,T_B)$

    \ForAll{memory block $b \in \textsc{SuspectBlocks}(O_{joint},P)$}
        \State $I^{solo}_{b} \gets \textsc{TraceFootprintInSoloTrace}(T_P,b)$
        \State $I^{multi}_{b} \gets \textsc{TraceFootprintInMergedTrace}(T_{merged},b)$
        \State $\Delta CP_b \gets \textsc{CalculateContentionPotentialGrowth}(I^{solo}_{b}, I^{multi}_{b}, \theta)$

        \If{$\Delta CP_b > 0$}
            \State $\mathcal{C}_P \gets \mathcal{C}_P \cup \{(b.address,\Delta CP_b,\textsc{VulnerableWindows}(b))\}$
        \EndIf
    \EndFor
\EndFor

\State $R_{final} \gets \textsc{GenerateDynamicRiskAssessment}(\{\mathcal{C}_A,\mathcal{C}_B\}, \theta)$
\State \Return $R_{final}$

\end{algorithmic}
\end{algorithm}

The final qualitative risk class is assigned by combining the static occupancy result with the temporal footprint result. A pair is classified as low risk when the combined footprint remains below the critical threshold. It is classified as low-to-medium risk when the footprint approaches the associativity without exceeding it. If one or more sets exceed the associativity, the pair is first
flagged as statically risky and then refined using $\Delta CP$. Larger $\Delta CP$ values indicate higher risk, whereas lower $\Delta CP$ values indicate medium risk because the spatial overlap is temporally dispersed.
\subsubsection{Step 1: Coarse-Grained Application Classification}

The first step classifies each application according to its non-stack access ratio. If both applications exhibit a very low proportion of non-stack accesses, the pair is classified as low risk because both memory footprints are expected to remain localized. If at least one application exhibits significant non-stack activity, the framework proceeds to cache-set occupancy analysis.

This first step is intentionally coarse-grained. It is designed to quickly eliminate clearly safe application pairs, but it does not replace set-level analysis. A pair containing one stack-dominated application and one data-intensive application may still be evaluated spatially to confirm whether any cache-set overlap exists.

\subsubsection{Step 2: Static Multicore Cache Profiling}

The second step evaluates the spatial distribution of the combined cache footprint. The framework merges the traces, computes joint cache-set occupancy, and generates a joint cache heatmap. If the combined number of unique memory blocks remains below the critical threshold $\theta$ for all cache sets, the pair is classified as low risk. If the combined footprint approaches the physical associativity but does not exceed it, the pair is classified as a near-capacity case. Such cases are conservatively marked as low-to-medium risk because unmodeled hardware behavior, such as pseudo-LRU replacement, prefetching, or speculative effects, may still trigger evictions.

If at least one cache set exceeds the hardware associativity $W$, the framework flags a static capacity violation. The corresponding cache sets and overload deltas are reported to the designer. These sets become candidates for temporal footprint analysis.

\subsubsection{Step 3: Temporal Footprint Analysis}

The third step refines the static assessment by identifying when cache conflicts are likely to occur. Static occupancy indicates where overload is possible, but it does not determine whether conflicting accesses occur within overlapping execution windows. Therefore, the framework computes intervening unique footprints over the interlaced trace and identifies vulnerable reuse windows.

The resulting $\Delta CP$ metric quantifies how much the co-running application increases the number of vulnerable windows compared with isolated execution. A high $\Delta CP$ value indicates that the co-runner significantly expands the target application's eviction-prone execution windows. A low $\Delta CP$ value indicates that the spatial overlap observed in the cache occupancy profile is temporally dispersed and may have limited impact on actual hardware behavior.

\subsection{Generated Reports}

The final output of the framework is an interference-awareness report. The report includes:
\begin{itemize}
    \item the stack and non-stack access ratio of each application;
    \item the cache-set occupancy profile of each application;
    \item the joint cache-set occupancy profile of the application pair;
    \item cache heatmaps showing access concentration and spatial distribution;
    \item the list of overloaded or near-capacity cache sets;
    \item vulnerable temporal windows identified by intervening-footprint analysis;
    \item the Contention Potential Growth ($\Delta CP$) values;
    \item the final qualitative risk classification.
\end{itemize}

The final objective of the report is not to provide a formal WCET bound. Instead, it provides an early indication of where interference may occur, which application pairings are most susceptible to shared-cache contention, and which hardware resources should be prioritized for later validation. This information can guide application placement, memory-layout decisions, workload selection, and hardware-validation experiments.

\section{Validation and Results}\label{results_section}

This section validates the proposed interference-awareness framework using benchmark application pairings executed on a real multicore embedded platform. The objective is not to demonstrate cycle-accurate prediction of execution time, but to evaluate whether the proposed profiling-based analysis can identify workload combinations that are likely to generate shared-cache interference.

The validation uses four benchmark pairings selected from the TACLeBench suite~\cite{tacleBench} and from custom memory- and cache-oriented workloads implementing classical algorithms. The selected pairs exercise different memory-access behaviors, ranging from highly data-intensive workloads to stack-dominated and low-contention cases. This diversity allows the evaluation of the progressive analysis workflow introduced in Section~\ref{subsec:risk_assessment}.

The evaluated configurations are:
\begin{enumerate}
    \item Dijkstra vs. Floyd-Warshall: a pair of highly data-intensive workloads expected to create broad cache-set pressure.
    \item FFT vs. Jacobi-2D: a near-capacity configuration where the combined cache occupancy approaches, but does not exceed, the physical associativity limit.
    \item AES-128 ECB vs. Conv2D: a pair that exhibits severe static cache-set occupancy, requiring temporal refinement to avoid over-pessimistic classification.
    \item Bsort vs. ADPCM Encoder: a pair with limited joint cache pressure, used to evaluate the framework's ability to identify low-risk co-scheduling candidates.
\end{enumerate}

The validation is organized around two complementary analyses. First, hardware performance counters are collected during isolated and co-running executions to provide a physical reference measurement of interference. Second, the same application pairs are analyzed using the proposed profiling framework to obtain non-stack access ratios, joint cache occupancy profiles, heatmaps, temporal footprint indicators, and final qualitative risk classifications. The hardware measurements are used only for validation; they are not required when applying the profiling framework during early design exploration.

\subsection{Hardware Reference Measurements}
\label{subsec:hardware_reference_measurements}

The hardware validation is performed on an NXP QorIQ P2020RDB-PCA evaluation board. The platform contains a dual-core e500v2 Power Architecture processor with a shared L2 cache. The applications are executed as bare-metal multicore projects, avoiding operating-system scheduling overhead and reducing noise caused by context switches.

For each benchmark pair, two execution scenarios are considered: isolated execution and co-running execution.

\subsubsection{Isolated Execution}

In the isolated scenario, the target application executes alone on one core while the second core remains inactive or executes no interfering workload. This execution provides the baseline behavior of the target application in the absence of intentional multicore interference.

During this baseline execution, the on-chip Performance Monitor Counters (PMCs) are used to measure total L2 data cache misses. This counter establishes a reference value for the target application's cache behavior. Other counters, such as execution cycles, memory-related events, or stall cycles, can be collected in the same way when supported by the target platform.

\subsubsection{Co-running Execution}

In the co-running scenario, the target application executes on one core while a second application executes concurrently on the other core. The co-running application generates additional pressure on shared resources such as the L2 cache, the memory bus, or the memory subsystem.

The same hardware counter is collected during co-running execution. The difference between isolated and co-running executions provides a hardware reference for the observed interference. For each measured counter $X$, the relative variation is computed as:
\begin{equation}
    \Delta X =
    \frac{X_{multi} - X_{single}}{X_{single}} \times 100 .
    \label{eq:counter_variation}
\end{equation}

A positive value of $\Delta X$ indicates that the corresponding event increased during co-running execution. In this study, the main counter is the total number of L2 data cache misses. Since the P2020RDB-PCA exposes this counter at the shared-cache level, the measurements report total L2 miss variation rather than per-core L2 miss attribution.

\subsection{Profiling-Based Interference-Awareness Results}
\label{subsec:profiling_results}

This subsection presents the profiling-based analysis for each application pair. The analysis follows the three-stage workflow defined in Section~\ref{subsec:risk_assessment}: coarse-grained non-stack screening, static joint cache profiling, and temporal footprint analysis when required.

\subsubsection{Case Study 1: Dijkstra vs. Floyd-Warshall}
\label{subsubsec:results_dijkstra_floyd}

The first case study evaluates the pairing of Dijkstra and Floyd-Warshall. The coarse-grained memory classification gives the following access distributions:
\begin{itemize}
    \item Dijkstra: Stack: 10,994,179 accesses; Non-stack: 2,037,674 accesses (15.6\%).
    \item Floyd-Warshall: Stack: 13,338,844 accesses; Non-stack: 1,607,493 accesses (10.8\%).
\end{itemize}

Both applications exceed the 10\% non-stack access threshold. Therefore, the pair cannot be filtered out by the coarse-grained screening step and must be analyzed using set-level cache profiling.

\begin{figure}[!t]
    \centering
    \includegraphics[width=0.8\linewidth]{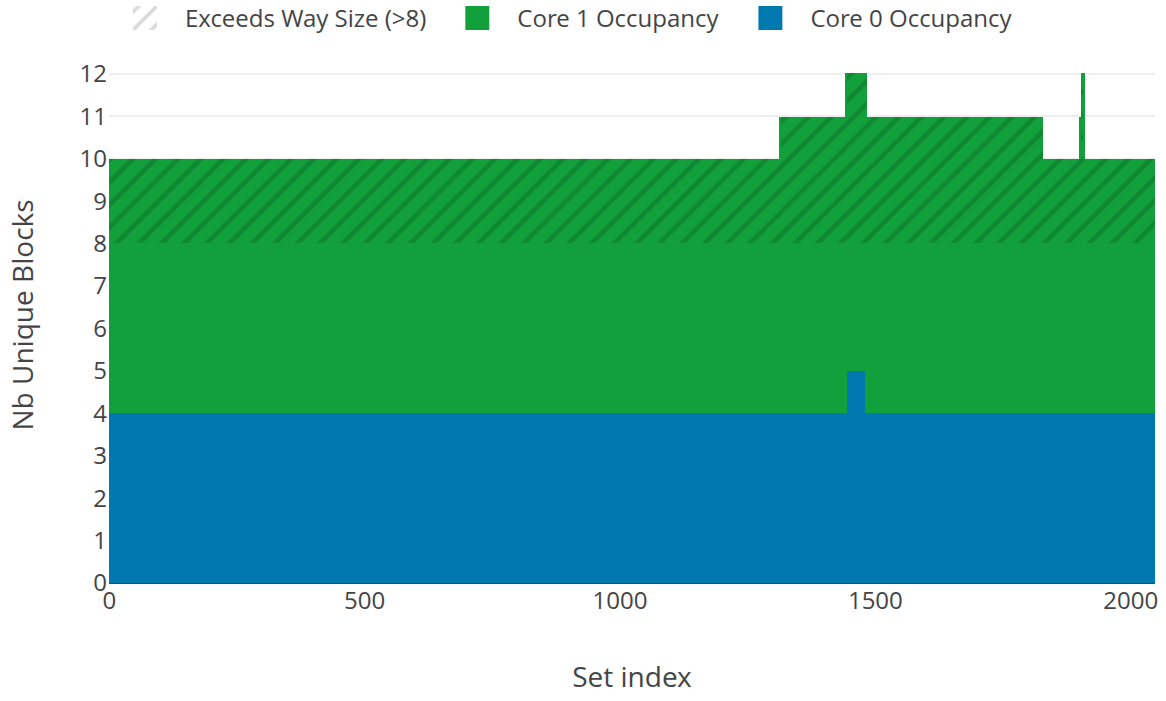}
    \caption{Joint cache occupancy profile for Dijkstra vs. Floyd-Warshall showing capacity violations across cache sets.}
    \label{fig:dijkstra_floyd_occupancy}
\end{figure}

Figure~\ref{fig:dijkstra_floyd_occupancy} shows the joint cache occupancy profile. The combined footprint reaches full cache-set utilization, and several sets exceed the 8-way hardware associativity by up to four additional unique blocks. This indicates a clear static capacity violation and identifies the pair as vulnerable to shared-cache eviction.

\begin{figure}[!t]
    \centering
    \includegraphics[width=0.8\linewidth]{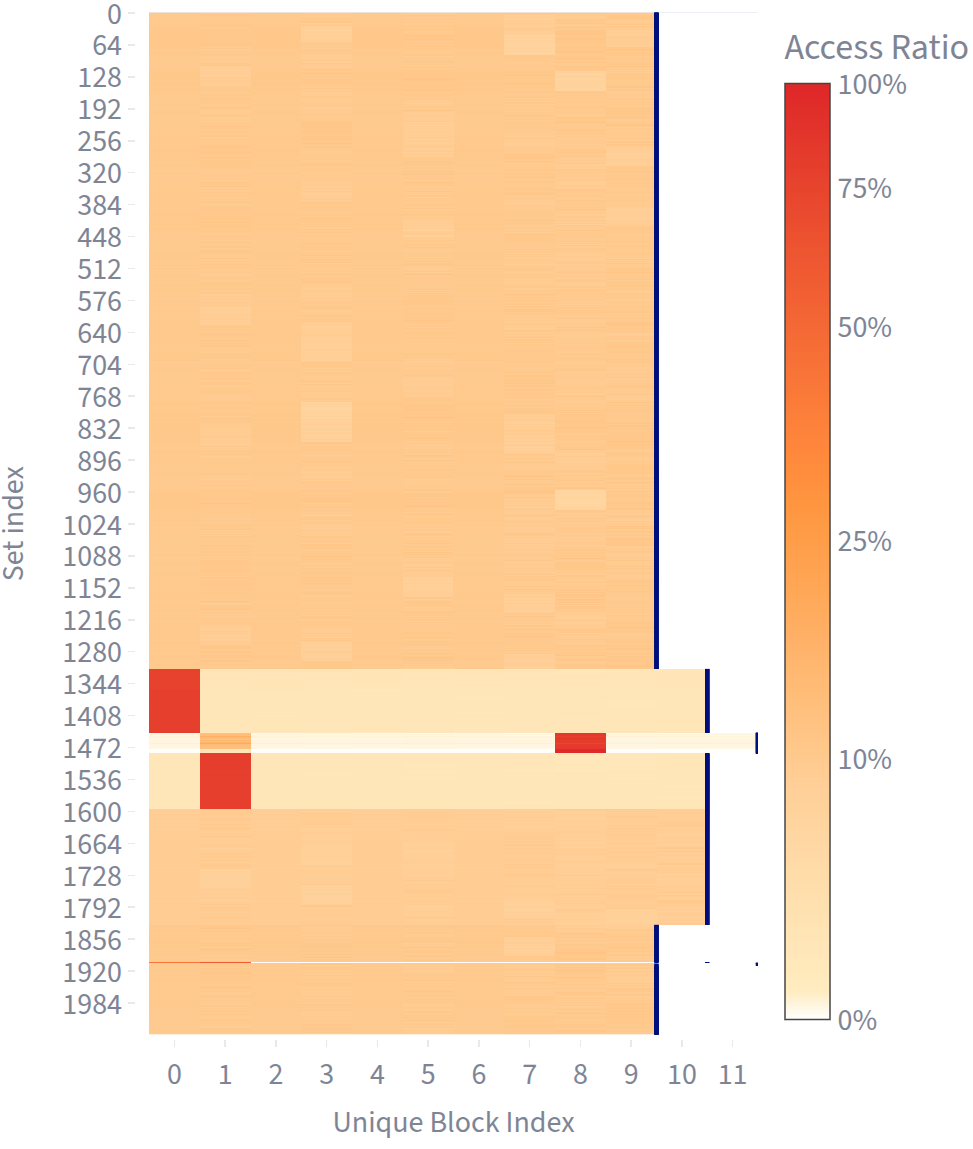}
    \caption{Joint L2 cache heatmap for Dijkstra vs. Floyd-Warshall illustrating spatial access distribution.}
    \label{fig:dijkstra_floyd_heat}
\end{figure}

Figure~\ref{fig:dijkstra_floyd_heat} shows the corresponding joint cache heatmap. Most cache sets exhibit a broadly distributed access pattern, meaning that many unique blocks compete for the available ways. A more concentrated access region appears between approximately sets 1300 and 1600, where specific blocks receive higher access density. Overall, the spatial distribution indicates sustained pressure on the cache replacement mechanism.

To refine the spatial analysis, the temporal footprint analysis computes Contention Potential Growth ($\Delta CP$) under different critical threshold values. Table~\ref{tab:dijkstra_floyd_cpg} reports the resulting values.

\begin{table}[!t]
\centering
\caption{Contention Potential Growth ($\Delta CP$) under varying threshold levels ($\beta$) for Dijkstra vs. Floyd-Warshall.}
\label{tab:dijkstra_floyd_cpg}
\scriptsize
\renewcommand{\arraystretch}{1.15}
\begin{tabular}{|c|c|}
\hline
\textbf{Critical Threshold ($\beta$)} & \textbf{Contention Potential Growth ($\Delta CP$)} \\
\hline
62.5\% (5 ways)  & 48.35\%  \\
\hline
75.0\% (6 ways)  & 55.85\%  \\
\hline
87.5\% (7 ways)  & 63.48\%  \\
\hline
100.0\% (8 ways) & 281.28\% \\
\hline
\end{tabular}
\end{table}

The $\Delta CP$ values increase as the critical threshold approaches the physical associativity limit. At $\beta=100\%$, the value reaches 281.28\%, indicating that co-execution substantially increases the number of vulnerable reuse windows. Based on the combination of non-stack activity, static capacity violations, broad heatmap pressure, and high temporal footprint growth, the framework classifies the Dijkstra/Floyd-Warshall pair as \textit{high risk}.

\subsubsection{Case Study 2: FFT vs. Jacobi-2D}
\label{subsubsec:results_fft_jacobi}

The second case study evaluates the pairing of FFT and Jacobi-2D. The initial classification yields:
\begin{itemize}
    \item FFT: Stack: 4,684,198 accesses; Non-stack: 1,097,212 accesses (18.98\%).
    \item Jacobi-2D: Stack: 7,997,712 accesses; Non-stack: 1,557,334 accesses (16.30\%).
\end{itemize}

Both applications exceed the 10\% non-stack threshold. Therefore, the pair is passed to the joint cache profiling stage.

\begin{figure}[!t]
    \centering
    \includegraphics[width=0.8\linewidth]{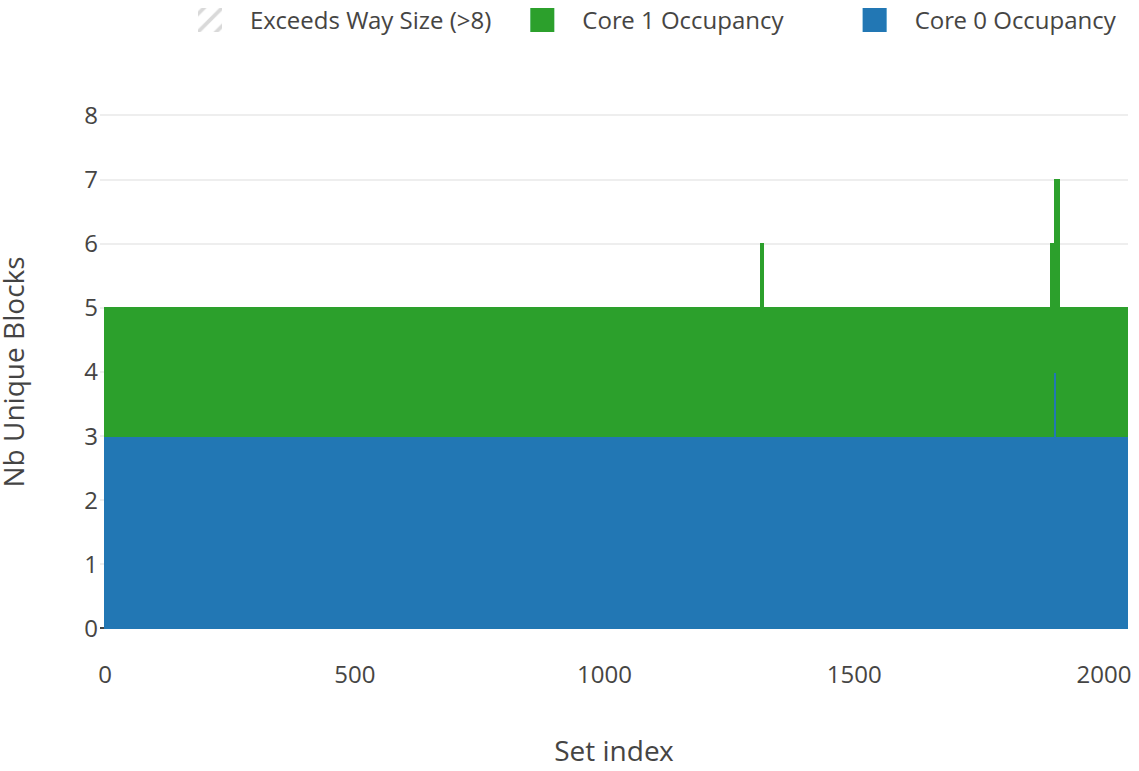}
    \caption{Joint cache occupancy profile for FFT vs. Jacobi-2D. Although the occupancy remains below the physical 8-way associativity, several sets approach near-capacity levels.}
    \label{fig:fft_jacobi_occupancy}
\end{figure}

Figure~\ref{fig:fft_jacobi_occupancy} shows the joint cache occupancy profile. The maximum combined occupancy reaches seven unique blocks in only a small number of sets, while some sets reach six unique blocks and most remain around five. Therefore, no cache set exceeds the 8-way physical associativity.

A purely static interpretation could classify this pair as safe because the hardware associativity is not exceeded. However, several sets operate close to the critical threshold. Since real hardware may use pseudo-LRU replacement, prefetching, speculative accesses, and other effects not modeled by the framework, near-capacity behavior is conservatively flagged. The framework therefore classifies this pair as \textit{low-to-medium risk}. This classification is intentionally conservative: it indicates that the pair should not be treated as a clear static violation, but it should still be considered for hardware validation.

\subsubsection{Case Study 3: AES-128 ECB vs. Conv2D}
\label{subsubsec:results_aes_conv}

The third case study evaluates the pairing of AES-128 ECB and Conv2D. The initial memory classification gives:
\begin{itemize}
    \item AES-128 ECB: Stack: 9,021,201 accesses; Non-stack: 1,487,529 accesses (14.16\%).
    \item Conv2D: Stack: 14,368,719 accesses; Non-stack: 2,082,867 accesses (12.66\%).
\end{itemize}

Both applications exceed the 10\% non-stack threshold, triggering detailed spatial and temporal analysis.

\begin{figure}[!t]
    \centering
    \includegraphics[width=0.8\linewidth]{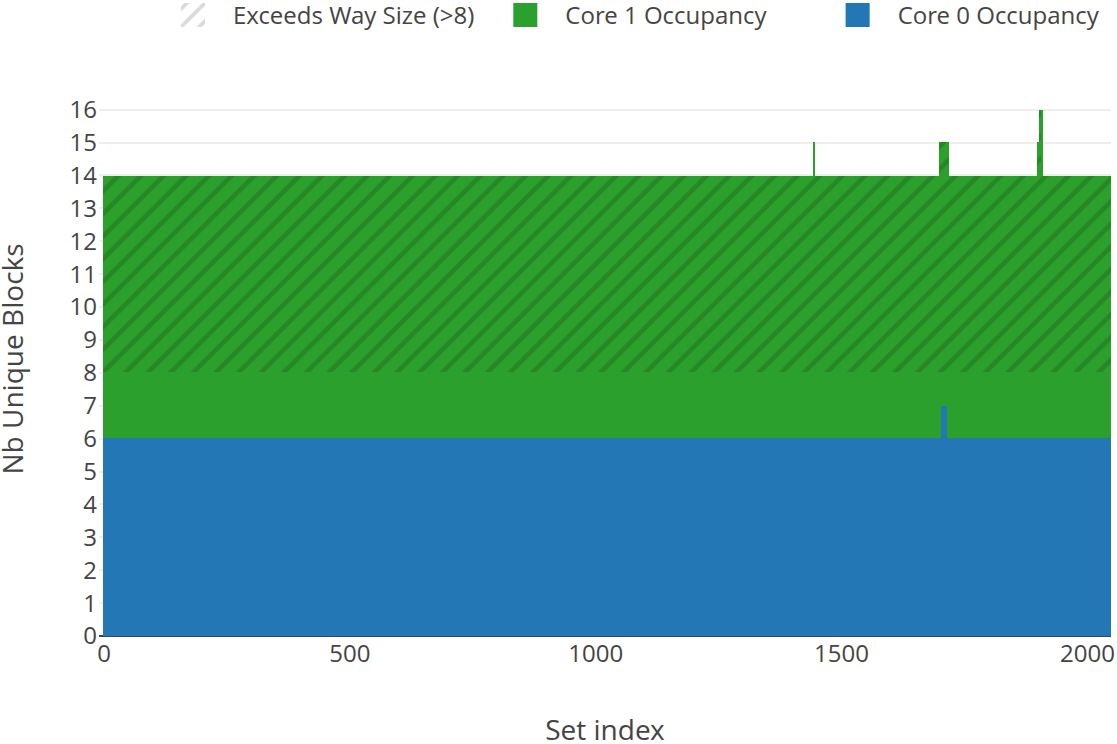}
    \caption{Joint cache occupancy profile for AES-128 ECB vs. Conv2D showing severe spatial capacity overload across sets.}
    \label{fig:aes_conv_occupancy}
\end{figure}

Figure~\ref{fig:aes_conv_occupancy} shows the joint cache occupancy profile. The spatial occupancy reaches up to 16 unique blocks per set, with many sets between 14 and 15 unique blocks. Since the target cache is 8-way set associative, this means that the combined footprint can exceed the physical associativity by six or more blocks. Based only on static occupancy, this pair would be classified as high risk.

\begin{figure}[!t]
    \centering
    \includegraphics[width=0.8\linewidth]{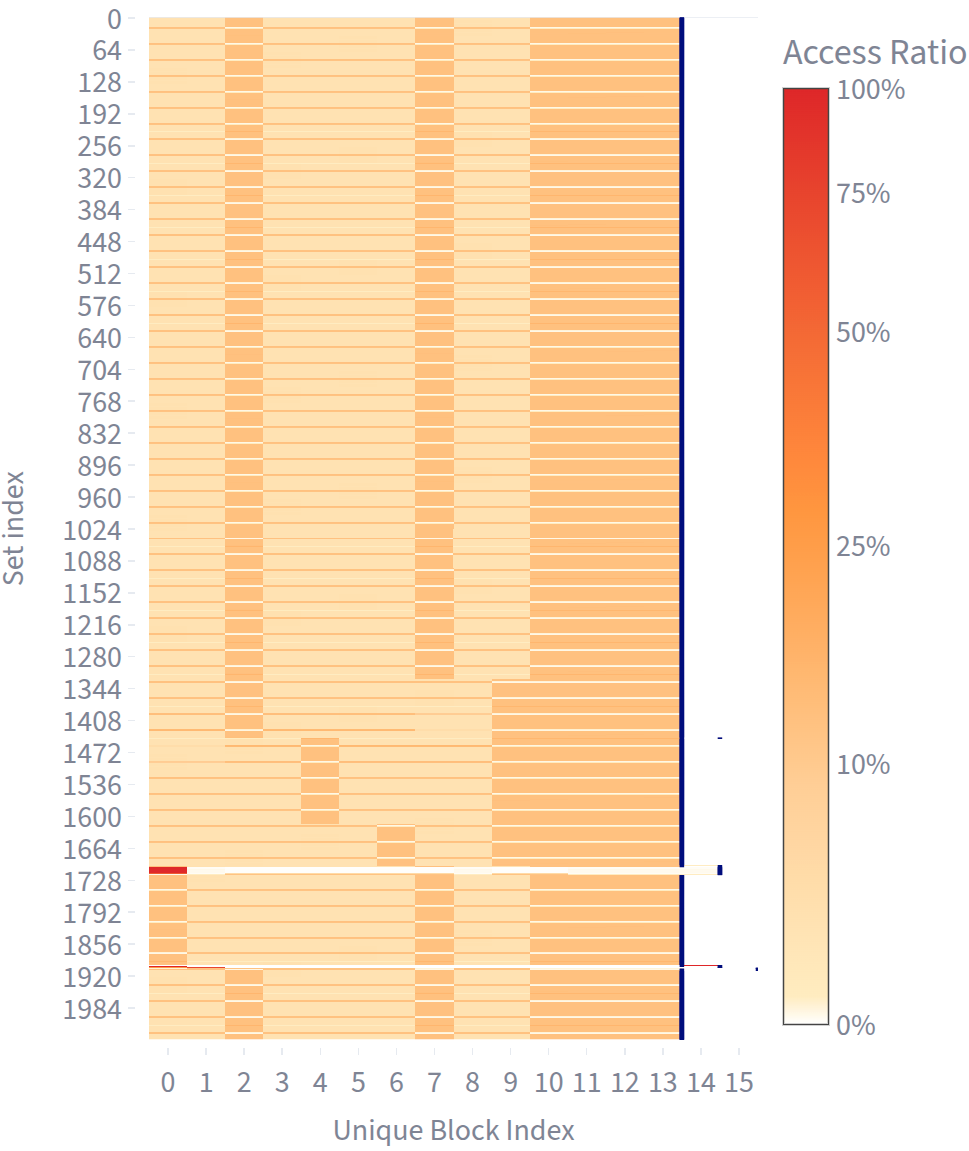}
    \caption{Joint cache heatmap for AES-128 ECB vs. Conv2D illustrating concentrated accesses to a subset of hot blocks.}
    \label{fig:aes_conv_heat}
\end{figure}

However, the heatmap in Figure~\ref{fig:aes_conv_heat} provides additional interpretation. Although many unique blocks map to the same sets, the access distribution is fairly uneven. A subset of blocks receives most of the accesses, while the remaining blocks are accessed less frequently. This suggests that the severe static footprint may overestimate the effective temporal eviction pressure.

To refine the risk classification, the framework applies temporal footprint analysis. Table~\ref{tab:aes_conv_cpg} reports the $\Delta CP$ values.

\begin{table}[!t]
\centering
\caption{Contention Potential Growth ($\Delta CP$) under varying threshold levels ($\beta$) for AES-128 ECB vs. Conv2D.}
\label{tab:aes_conv_cpg}
\scriptsize
\renewcommand{\arraystretch}{1.15}
\begin{tabular}{|c|c|}
\hline
\textbf{Critical Threshold ($\beta$)} & \textbf{Contention Potential Growth ($\Delta CP$)} \\
\hline
62.5\% (5 ways)  & 10.36\% \\
\hline
75.0\% (6 ways)  & 13.12\% \\
\hline
87.5\% (7 ways)  & 13.84\% \\
\hline
100.0\% (8 ways) & 78.58\% \\
\hline
\end{tabular}
\end{table}

Despite the severe static footprint, $\Delta CP$ remains below 14\% for thresholds up to 87.5\% of the cache capacity and reaches 78.58\% at the full associativity threshold. This indicates that the potentially conflicting accesses are dispersed over time and do not continuously expand vulnerable reuse windows. Consequently, the temporal analysis refines the initial high-risk static classification to a \textit{medium-risk} assessment.

\subsubsection{Case Study 4: Bsort vs. ADPCM Encoder}
\label{subsubsec:results_bsort_adpcm}

The final case study evaluates Bsort with ADPCM Encoder. The initial memory classification gives:
\begin{itemize}
    \item Bsort: Stack: 108,892 accesses; Non-stack: 30,388 accesses (21.82\%).
    \item ADPCM Encoder: Stack: 168,439 accesses; Non-stack: 674 accesses (0.40\%).
\end{itemize}

The ADPCM Encoder is almost entirely stack-dominated, with only 0.40\% non-stack accesses. However, because Bsort exceeds the 10\% non-stack threshold, the pair is not discarded solely by the first screening step. Instead, the framework proceeds to joint cache occupancy analysis to verify whether Bsort's data activity overlaps with any significant footprint from the co-running application.

\begin{figure}[!t]
    \centering
    \includegraphics[width=0.8\linewidth]{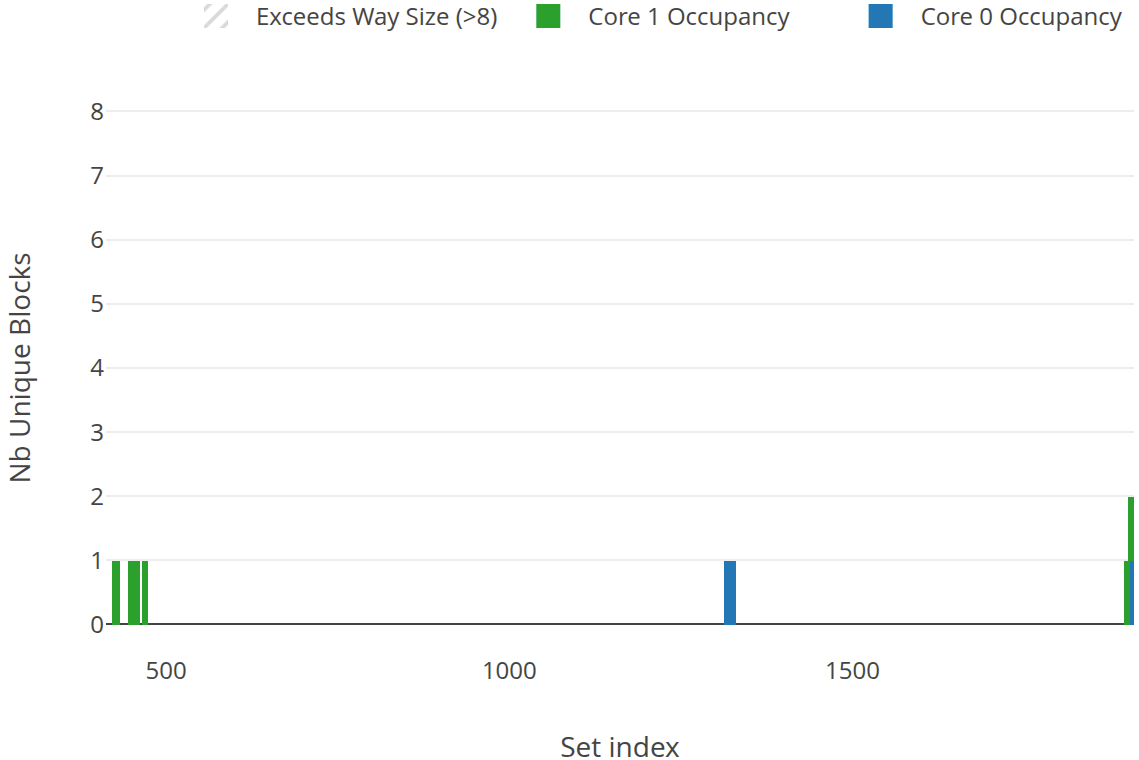}
    \caption{Joint cache occupancy profile for Bsort vs. ADPCM Encoder showing minimal set-level pressure, with occupancy peaking at only two unique blocks.}
    \label{fig:bsort_adpcm_occupancy}
\end{figure}

Figure~\ref{fig:bsort_adpcm_occupancy} shows the joint cache occupancy profile. The maximum combined occupancy across all cache sets is only two unique memory blocks. This is far below both the 8-way physical associativity and the default critical threshold used for conservative analysis.

Because the spatial pressure is minimal, the pair does not require temporal footprint analysis. The framework classifies Bsort/ADPCM Encoder as \textit{low risk}. This case demonstrates that the joint occupancy step can confirm safe co-scheduling candidates even when one application has a non-negligible non-stack access ratio.

\subsection{Hardware Validation Summary}
\label{subsec:hardware_validation_summary}

Table~\ref{tab:hardware_validation} summarizes the hardware measurements and compares them with the qualitative risk classifications produced by the framework.

\begin{table*}[]
\centering
\caption{Framework risk assessments compared with physical L2 cache-miss measurements on the P2020 platform.}
\label{tab:hardware_validation}
\scriptsize
\renewcommand{\arraystretch}{1.15}
\begin{tabular}{|l|c|c|c|c|}
\hline
\textbf{Application Pairing} &
\textbf{Baseline L2 Misses} &
\textbf{Co-run L2 Misses} &
\textbf{L2 Miss Variation} &
\textbf{Predicted Risk} \\
\hline
Dijkstra vs. Floyd-Warshall &
53,920 &
168,819 &
+213.09\% &
High \\
\hline
FFT vs. Jacobi-2D &
430,942 &
428,341 &
-0.60\% &
Low-to-medium / Conservative \\
\hline
AES-128 ECB vs. Conv2D &
261,216 &
271,158 &
+3.81\% &
Medium \\
\hline
Bsort vs. ADPCM Encoder &
120 &
118 &
-1.67\% &
Low \\
\hline
\end{tabular}
\end{table*}

The hardware measurements are consistent with the intended role of the framework as an early interference-awareness tool.

For Dijkstra vs. Floyd-Warshall, the framework predicts high risk due to non-stack activity, static cache-set overload, broad heatmap pressure, and high $\Delta CP$. The hardware counters confirm this prediction with a 213.09\% increase in L2 data cache misses.

For FFT vs. Jacobi-2D, the framework produces a conservative low-to-medium warning because several cache sets approach the associativity limit without exceeding it. The hardware measurement shows a negligible variation of -0.60\%. This case illustrates a safe false positive: the framework intentionally warns about near-capacity configurations even when the measured execution does not exhibit significant cache degradation.

For AES-128 ECB vs. Conv2D, the static occupancy profile alone suggests high risk, but the heatmap and $\Delta CP$ analysis refine the classification to medium risk. The hardware result shows a modest 3.81\% increase in L2 misses, supporting the temporal refinement.

For Bsort vs. ADPCM Encoder, the framework predicts low risk after joint cache occupancy confirms minimal set-level pressure. The hardware measurement also shows negligible variation, with only two fewer L2 misses during co-running execution. This validates the ability of the framework to identify clearly low-risk co-scheduling candidates without executing the more expensive temporal analysis.

Overall, the validation shows that the proposed framework identifies severe cache-interference cases, provides conservative warnings for near-capacity cases, refines over-pessimistic static classifications through temporal footprint analysis, and filters low-risk pairings when joint cache pressure is minimal.
\section{Discussion}\label{discussion_section}

This section discusses the interpretation of the validation results, the relationship between the profiling-based risk indicators and the hardware measurements, and the main limitations of the current framework. The objective of the proposed approach is not to replace cycle-accurate simulation, formal WCET analysis, or exhaustive hardware testing. Instead, the framework is intended to provide an early and interpretable interference-awareness layer that helps designers identify risky application pairings and prioritize hardware validation.

\subsection{Interpretation of the Validation Results}
\label{subsec:discussion_validation_interpretation}

The validation results show that memory-access profiling provides useful information for estimating cache-related multicore interference tendencies. The four evaluated application pairings illustrate different situations that may occur during multicore integration.

The Dijkstra/Floyd-Warshall pair represents a clear high-risk case. Both applications exhibit significant non-stack memory activity, the joint cache occupancy exceeds the 8-way associativity in several sets, and the temporal footprint analysis shows a strong increase in Contention Potential Growth ($\Delta CP$). This prediction is consistent with the hardware measurements, where the total L2 data cache misses increase by 213.09\%. This result confirms that the proposed spatial and temporal indicators can identify severe shared-cache contention.

The Bsort/ADPCM Encoder pair represents a clear low-risk case. Although Bsort exhibits a non-negligible non-stack access ratio, ADPCM Encoder is almost entirely stack dominated. More importantly, the joint cache occupancy reaches only two unique blocks per set, which is far below both the physical associativity and the conservative critical threshold. The hardware measurements confirm this low-risk prediction, with negligible L2 cache-miss variation. This case shows that the framework can identify safe co-scheduling candidates without requiring detailed temporal footprint analysis.

The AES-128 ECB/Conv2D pair illustrates the importance of temporal refinement. The static joint occupancy profile alone suggests severe cache pressure because the number of unique blocks per set can reach twice the physical associativity. However, the heatmap and $\Delta CP$ analysis show that the accesses are not uniformly conflicting over time. The temporal footprint analysis therefore refines the initial static warning from high risk to medium risk. The hardware measurements support this refinement, showing only a modest 3.81\% increase in L2 cache misses.

The FFT/Jacobi-2D pair demonstrates the conservative nature of the framework. The joint occupancy remains below the physical associativity, but several sets approach near-capacity levels. The framework therefore classifies the pair as low-to-medium risk. The hardware measurements show negligible L2 miss variation, indicating that no significant L2 cache-miss increase was observed in the tested execution. This is a safe false positive rather than a false negative. In early-stage analysis for embedded or safety-conscious systems, such conservatism is preferable because it encourages validation of near-capacity configurations instead of incorrectly dismissing them as risk-free.

\subsection{Relationship Between Estimated Risk and Hardware Measurements}
\label{subsec:discussion_correlation}

The comparison with hardware performance counters should be interpreted qualitatively rather than as an exact numerical prediction. The proposed framework estimates interference potential from source-level memory-operation traces, cache-set mappings, joint occupancy, heatmaps, and temporal footprint growth. In contrast, the hardware counter reports the final effect of the actual execution on the target platform.

In this paper, the main validation counter is the total number of L2 data cache misses. This counter is relevant because the proposed framework focuses primarily on shared-cache interference. However, the P2020RDB-PCA platform reports the considered L2 miss count at the shared-cache level rather than as a per-core attribution. Therefore, the measurement indicates that the co-running execution changes the shared-cache behavior, but it does not directly identify which core or which memory block caused each miss.

Moreover, hardware measurements capture the combined effect of several microarchitectural mechanisms. Cache replacement policy, prefetching, write buffers, bus arbitration, memory-controller behavior, and instruction-level effects may all influence the measured L2 miss count. Consequently, the goal is not to obtain a perfect match between $\Delta CP$ and the measured miss variation. Instead, the relevant question is whether the framework correctly distinguishes high-risk, medium-risk, conservative-warning, and low-risk pairings. The validation results indicate that it does.

\subsection{Role of Static Occupancy, Heatmaps, and Temporal Footprint Analysis}
\label{subsec:discussion_static_temporal}

The results confirm that no single indicator is sufficient to characterize multicore cache interference. The non-stack access ratio is useful as a coarse screening criterion, but it cannot determine whether two applications actually compete for the same cache sets. Joint cache occupancy provides a stronger spatial indicator by showing whether the combined footprint exceeds or approaches the cache associativity. However, occupancy alone may be pessimistic because it ignores when the accesses occur.

The cache heatmaps provide an interpretable bridge between spatial occupancy and temporal behavior. They show whether accesses are broadly distributed across many blocks or concentrated on a small subset of frequently reused blocks. This distinction is important because two workloads may have similar occupancy values but different effective interference behavior. A broad and uniform access distribution tends to exert sustained pressure on the replacement mechanism, whereas a skewed distribution may produce less interference if most of the unique blocks are rarely accessed.

Temporal footprint analysis further refines the assessment by considering reuse windows. By tracking the intervening unique blocks between consecutive accesses to the same target block, the framework estimates whether the co-running application expands eviction-vulnerable windows. The $\Delta CP$ metric therefore helps distinguish between static overload that is temporally harmful and static overload that is dispersed across execution phases. This explains why AES-128 ECB/Conv2D is refined from a high static risk to a medium final risk.

\subsection{Safety-Oriented Conservatism}
\label{subsec:discussion_conservatism}

The framework intentionally favors conservative warnings over aggressive filtering. In early-stage multicore integration, a false negative is more problematic than a false positive. A false negative would incorrectly classify a risky application pair as safe, potentially allowing harmful co-scheduling decisions to pass without hardware validation. A false positive, by contrast, only causes an application pair to be prioritized for further analysis or measurement.

The FFT/Jacobi-2D case illustrates this design choice. The hardware result shows negligible L2 cache-miss variation, but the framework still marks the pair as low-to-medium risk because some sets approach the physical associativity. This warning is conservative, but it is justified by the fact that the framework does not model all hardware effects that may trigger early eviction, such as pseudo-LRU behavior, prefetching, or speculative memory activity. In this sense, the framework is better understood as a screening and prioritization tool rather than a final certification mechanism.

\subsection{Practical Use During Multicore Integration}
\label{subsec:discussion_practical_use}

The proposed framework can support multicore integration in several ways. First, it can help designers identify application pairings that should not be co-scheduled without further hardware validation. Second, it can highlight overloaded or near-capacity cache sets, providing clues about where memory-layout changes may reduce interference. Third, it can reduce the number of hardware experiments by filtering clearly low-risk configurations and focusing attention on high-risk or ambiguous cases.

The generated reports are also useful because they are interpretable. Instead of only reporting that a pair is risky, the framework explains whether the risk comes from non-stack memory activity, static cache-set overload, broad heatmap pressure, or temporal growth of vulnerable reuse windows. This interpretability is important for embedded-system designers because it can guide concrete actions such as changing core allocation, modifying data layout, adjusting memory placement, or selecting which workload combinations to validate first on hardware.

\subsection{Limitations}
\label{subsec:discussion_limitations}

The current framework has several limitations that should be considered when interpreting the results.

\subsubsection{Source-Level Trace Approximation}

The generated trace is a source-level approximation of memory behavior. It captures memory-operation patterns through type-based profiling, but it does not reproduce the exact binary-level load and store stream generated by the compiler. Compiler optimizations, register allocation, instruction scheduling, inlining, and library implementations may change the final memory behavior observed on hardware. Therefore, the trace should be interpreted as an approximation of memory-access tendencies rather than an exact execution trace.

\subsubsection{Coverage of the Profiling Mechanism}

The type-based profiling mechanism reduces manual instrumentation, but it may not automatically capture all memory interactions. Pointer arithmetic, direct array indexing, structure field accesses, inline assembly, optimized library calls, and explicit memory functions such as \texttt{memcpy} and \texttt{memset} may require additional wrappers or compiler-assisted instrumentation. Missing events may lead to underestimation of cache pressure or memory intensity. For this reason, trace completeness should be checked carefully when applying the framework to complex C applications.

\subsubsection{No Cycle-Accurate Cache Simulation}

The framework does not simulate exact cache replacement behavior. It estimates cache pressure by mapping memory blocks to cache sets and by computing occupancy and intervening unique footprints. This approximation is useful for early awareness, but it does not determine exactly which cache line is present in the cache at each cycle. Replacement policies such as LRU, pseudo-LRU, random replacement, or implementation-specific policies may lead to different eviction behavior.

\subsubsection{Simplified Co-execution Approximation}

The interlaced trace provides an approximation of co-execution, not an exact reproduction of multicore timing. If precise timestamps are unavailable, the framework must rely on an ordering policy to merge traces. This may overestimate or underestimate temporal overlap depending on the actual execution schedule, memory latency, and synchronization behavior. Window-based and phase-based extensions can reduce this limitation by analyzing smaller execution intervals.

\subsubsection{Cache Coherency Not Explicitly Modeled}

The current implementation does not explicitly model cache-coherency protocols or snooping traffic. This assumption is reasonable for the evaluated scenarios because the applications are considered independent and do not intentionally share writable data. However, if applications share memory, writes from one core may invalidate or update cache lines used by another core. In such cases, coherence-induced traffic and ownership transfers may become significant sources of interference.

\subsubsection{Hardware-Dependent Effects}

The framework relies on architectural parameters such as cache size, associativity, cache-line size, and address mapping. If these parameters are incomplete or inaccurate, the resulting estimates may not reflect the actual hardware behavior. In addition, undocumented mechanisms such as prefetchers, write buffers, internal arbitration policies, and memory-controller scheduling may influence the measured behavior but are not fully captured by the current model.

\subsubsection{Limited Hardware Counter Attribution}

The validation uses total L2 data cache misses as the main hardware reference. This counter is appropriate for evaluating cache-related interference, but it does not provide complete attribution. In particular, it does not identify which application, memory address, or cache set caused each miss. Additional counters, address tracing, or platform-specific monitoring features would improve the resolution of the validation.

\subsection{Future Work}
\label{subsec:discussion_future_work}

Several extensions can improve the precision and applicability of the framework.

First, a trace-driven cache simulator could be integrated as an optional detailed analysis mode. Such a simulator would model replacement policies and estimate actual hits, misses, and evictions more explicitly. To preserve the lightweight nature of the framework, this mode could be applied only to application pairings already flagged as suspicious by the fast analysis.

Second, future work could extend the framework toward coherency-aware interference estimation. This would require detecting shared memory regions, identifying read/write sharing patterns, and estimating invalidation or ownership-transfer effects. Such an extension would make the framework applicable to producer-consumer workloads, shared buffers, and lock-based multicore applications.

Third, the temporal analysis could be extended with window-based and phase-based profiling. Instead of computing one global interference indicator over the complete trace, the framework could compute occupancy and $\Delta CP$ over execution windows. This would help identify when interference risk is highest and reduce overestimation caused by temporally distant accesses.

Fourth, automatic workload classification could be improved using threshold-based rules or clustering techniques. Applications could be classified according to non-stack access ratio, cache-set concentration, heatmap distribution, and temporal footprint growth. This would make the framework more scalable for larger benchmark suites and industrial software collections.

Finally, data-driven calibration could be introduced when sufficient hardware-validation data are available. Regression models or supervised learning techniques could learn the relationship between profiling-derived indicators and measured cache-miss or execution-time variation on a given platform. Such models should complement, not replace, the interpretable metrics used by the framework. %optional
\section{Conclusion}\label{conclusion_section}

This paper presented a profiling-based framework for early multicore interference awareness in embedded systems. The work was motivated by the fact that applications executing on different cores may still interact through shared hardware resources, even when they are logically independent at the software level. Such interactions can lead to cache eviction, memory-path contention, and execution-time variability, making multicore integration difficult to assess before deployment.

The proposed framework combines source-level memory-operation profiling with hardware-aware trace analysis. Instead of relying on cycle-accurate simulation, formal WCET bounding, or exhaustive co-execution experiments, the framework uses application memory traces and architectural parameters to approximate how workloads use shared cache resources. The analysis maps memory operations to cache sets and memory blocks, generates cache-occupancy profiles and heatmaps, and provides interpretable information about spatial cache pressure.

A progressive three-stage workflow was introduced. First, a coarse-grained screening step classifies applications using the ratio of non-stack memory accesses. This step helps identify workload pairs with limited potential for scattered memory pressure. Second, a static multicore cache-profiling step merges the traces of candidate co-running applications and evaluates whether their combined unique-block footprint exceeds or approaches the cache associativity. Third, a temporal footprint analysis tracks intervening unique blocks between consecutive accesses to the same memory block. This analysis computes Contention Potential Growth ($\Delta CP$), which quantifies the additional eviction-vulnerable reuse windows introduced by a co-running application.

The framework was validated using four application pairings executed on an NXP QorIQ P2020RDB-PCA dual-core platform. Hardware performance counters measuring total L2 data cache misses were collected during isolated and co-running executions and used as validation references. The Dijkstra/Floyd-Warshall pair was classified as high risk, which was confirmed by a 213.09\% increase in L2 cache misses. The Bsort/ADPCM Encoder pair was classified as low risk, consistent with negligible cache-miss variation. The AES-128 ECB/Conv2D pair demonstrated the value of temporal refinement: although static occupancy suggested severe overload, the $\Delta CP$ analysis refined the classification to medium risk, consistent with the modest 3.81\% increase in L2 misses. The FFT/Jacobi-2D pair illustrated the conservative behavior of the framework, producing a low-to-medium warning despite negligible measured degradation.

Overall, the results show that memory-access profiling can provide practical and interpretable early indicators of cache-related multicore interference. The framework does not provide formal timing guarantees and does not replace hardware validation or cycle-accurate cache simulation. Instead, it serves as an intermediate analysis layer between purely analytical models and physical measurements. It helps designers identify risky workload pairings, overloaded or near-capacity cache sets, and execution windows vulnerable to cross-core cache eviction.

Several limitations remain. The current implementation relies on source-level trace approximation, does not explicitly model cache-coherency traffic, and does not simulate detailed replacement policies, prefetching, write buffers, or memory-controller scheduling. Its accuracy also depends on trace representativeness and on the completeness of the profiling mechanism, especially for pointers, arrays, structures, library calls, and compiler-optimized code. Future work could extend the framework with optional trace-driven cache simulation, coherency-aware analysis, window-based and phase-based profiling, automatic workload classification, and data-driven calibration using larger sets of hardware performance-counter measurements.

%IEEE STYLE REFERENCES
\bibliographystyle{ieeetr}
%ACM STYLE REFERENCES
%\bibliographystyle{ACM-Reference-Format}

\bibliography{mybib}
\end{document}